\documentclass[a4paper,10pt]{article}
\usepackage{stmaryrd}
\usepackage{amsfonts}
\usepackage{bbm}
\usepackage{amscd}
\usepackage{mathrsfs}
\usepackage{latexsym,amssymb,amsmath,amscd,amscd,amsthm,amsxtra,xypic}
\usepackage[dvips]{graphicx}
\usepackage[utf8]{inputenc}
\usepackage[T1]{fontenc}
\usepackage{lmodern}
\usepackage{amssymb}
\usepackage[all]{xy}
\usepackage{nicefrac,mathtools,enumitem}
\usepackage{microtype}

\newtheorem{thm}{Theorem}[section]
\newtheorem{lem}[thm]{Lemma}
\newtheorem{cor}[thm]{Corollary}
\newtheorem{pro}[thm]{Proposition}
\newtheorem{ex}[thm]{Example}
\newtheorem{rmk}[thm]{Remark}
\newtheorem{defi}[thm]{Definition}

\newcommand{\be }{\begin{equation}}
\newcommand{\ee }{\end{equation}}

\newcommand{\pf}{\noindent{\bf Proof.}\ }

\newcommand{\g}{\frkg}

\newcommand{\huaA}{\mathcal{A}}

\newcommand{\huaF}{\mathcal{F}}
\newcommand{\huaG}{\mathcal{G}}

\newcommand{\huaV}{\mathcal{V}}

\newcommand{\frkd}{\mathfrak d}
\newcommand{\frke}{\mathfrak e}

\newcommand{\frkg}{\mathfrak g}
\newcommand{\frkh}{\mathfrak h}

\newcommand{\frkl}{\mathfrak l}

\newcommand{\frkr}{\mathfrak r}

\def\qed{\hfill ~\vrule height6pt width6pt depth0pt}

\newcommand{\Courant}[1]{\left\llbracket  #1\right\rrbracket }

\newcommand{\br}[1]{   [ \cdot,    \cdot  ]   }
\newcommand{\id}{\mathrm{id}}
\newcommand{\inc}{\mathbbm{i}}

\newcommand{\dM}{\mathrm{d}}

\newcommand{\Hom}{\mathrm{Hom}}
\newcommand{\Der}{\mathrm{Der}}

\newcommand{\gl}{\mathfrak {gl}}

\newcommand{\Ker}{\mathrm{Ker}}

\newcommand{\End}{\mathrm{End}}
\newcommand{\ad}{\mathrm{ad}}

\newcommand{\ve}{\mathrm{v}}
\newcommand{\Vect}{\mathrm{Vect}}

\newcommand{\V}{\mathbb{V}}
\newcommand{\A}{\mathbb{A}}

\begin{document}
\title{
{Categorification of Pre-Lie Algebras and Solutions of 2-graded Classical Yang-Baxter Equations
\thanks
 {
Research supported by NSFC (11101179).
 }
} }
\author{
Yunhe Sheng \\
Department of Mathematics, Jilin University,\\
 Changchun 130012, Jilin, China
\\\vspace{3mm}
email: shengyh@jlu.edu.cn}
\date{}
\footnotetext{{\it{Keyword}: pre-Lie $2$-algebras, pre-Lie$_\infty$-algebras, Lie $2$-algebras, $\mathcal O$-operators, $2$-graded classical Yang-Baxter equations }} \footnotetext{{\it{MSC}}: 17B99, 55U15.}
\maketitle

\begin{abstract}
In this paper, we introduce the notion of a pre-Lie 2-algebra, which is a categorification of a pre-Lie algebra. We prove that the category of pre-Lie 2-algebras and the category of 2-term pre-Lie$_\infty$-algebras are equivalent. We classify skeletal pre-Lie 2-algebras by the third cohomology of a pre-Lie algebra. We  prove that crossed modules of pre-Lie algebras are in one-to-one correspondence with strict pre-Lie 2-algebras. $\mathcal O$-operators on Lie 2-algebras are introduced, which can be used to construct pre-Lie 2-algebras. As an application, we give solutions of 2-graded classical Yang-Baxter equations in some semidirect product Lie 2-algebras.
\end{abstract}


\section{Introduction}

Pre-Lie
algebras (or left-symmetric algebras, Vinberg algebras, and etc.) arose from the study of affine
manifolds and affine Lie groups, convex homogeneous cones and
deformations of associative algebras. They appeared in many fields
in mathematics and mathematical physics (see the survey article
\cite{leftsymm4} and the references therein).  The beauty of a pre-Lie algebra is that the
commutator gives rise to a Lie algebra and the left multiplication
gives rise to a representation of the commutator Lie algebra. So
pre-Lie algebras naturally play important roles in the study
involving the representations of Lie algebras  on the underlying
spaces of the Lie algebras themselves or their dual spaces. For
example, they are the underlying algebraic structures of the
non-abelian phase spaces of Lie algebras
\cite{non-abelian phase spaces,Kupershmidt1}, which lead to a
bialgebra theory of pre-Lie algebras \cite{Left-symmetric
bialgebras}. They are also regarded as the algebraic structures
``behind'' the classical Yang-Baxter equations (CYBE) and they provide a
construction  of solutions of CYBE in
certain semidirect product Lie algebras (that is, over the
``double'' spaces) induced by pre-Lie algebras
\cite{Bai:CYBE,Kupershmidt2}.
Furthermore, pre-Lie
algebras are also regarded as the underlying algebraic structures of
symplectic Lie algebras \cite{chu}, which coincides with
 Drinfeld's  observation of the correspondence between the invertible (skew-symmetric) classical $r$-matrices and the symplectic forms
 on Lie algebras \cite{D}. In \cite{chapoton}, the authors studied pre-Lie algebras using the theory of operads, and introduced the notion of a pre-Lie$_\infty$-algebra. The author also proved that the PreLie operad is Koszul.  The PreLie operad is further studied in \cite{chapotonoperad} recently.

 $\mathcal O$-operators on a Lie algebra $\g$ associated to a representation $(V;\rho)$ were introduced in \cite{Kupershmidt2} inspired by the study of the operator form of the CYBE. See \cite{STS} for more details. On one hand, an $\mathcal O$-operator could give rise to a pre-Lie algebra structure on $V$. On the other hand, an $\mathcal O$-operator could give rise to a solution of the CYBE in the semidirect product Lie algebra $\g\ltimes_{\rho^*}V$.

 Recently, people have paid more attention to higher categorical
structures with motivations from string theory. One way to provide
higher categorical structures is by categorifying existing
mathematical concepts. One of the simplest higher structures is a
$2$-vector space, which is a categorified  vector space. If we
further put Lie algebra structures on $2$-vector spaces, then we
obtain  Lie $2$-algebras \cite{baez:2algebras}. The
Jacobi identity is replaced by a natural transformation, called the
Jacobiator, which also satisfies some coherence laws of its own. It is well-known that the category of Lie 2-algebras is equivalent to the category of  2-term
$L_\infty$-algebras
\cite{baez:2algebras}. The concept of an $L_\infty$-algebra (sometimes
called a strongly homotopy (sh) Lie algebra)  was originally
introduced in \cite{stasheff:introductionSHLA,stasheff:shla} as a
model for ``Lie algebras that satisfy the Jacobi identity up to all
higher homotopies''. The structure of a Lie 2-algebra appears in many areas such as
string theory \cite{baez:string}, higher symplectic geometry
\cite{baez:classicalstring}, and Courant algebroids \cite{Roytenberg1}.

The first aim of this paper is to category the relation between $\mathcal O$-operators, pre-Lie algebras and Lie algebras. We introduce the notion of an $\mathcal O$-operator on a Lie 2-algebra associated to a representation and the notion of a pre-Lie 2-algebra, and establish the following commutative diagram:
         $$
\xymatrix{
\mathcal O\mbox{-operators~ on~ Lie~ 2-algebras}\ar[r]^{}&\mbox{pre-Lie~ 2-algebras}
\ar[r]^{}& \mbox{Lie~2-algebras}\\
  \mathcal O\mbox{-operators~ on~ Lie~ algebras}\ar[u]_{\mbox{categorification}}\ar[r]&\mbox{pre-Lie~algebras}\ar[u]_{\mbox{categorification}}\ar[r]& \mbox{Lie~algebras.}\ar[u]_{\mbox{categorification}}}
                $$

In \cite{BaiShengZhu}, the authors  introduced the notion of an $L_\infty[l,k]$-bialgebra. In particular, an $L_\infty[0,1]$-bialgebra is a Lie 2-bialgebra, which is a certain categorification of the concept of a Lie
bialgebra. See \cite{chenweak,olga,Merkulov2} for more details along this direction. In \cite{chen2group}, the authors integrated a Lie 2-bialgebra to a quasi-Poisson 2-group, which generalizes the fact that a Lie bialgebra could be integrated to a Poisson-Lie group.
2-graded classical Yang-Baxter equations were established in \cite{BaiShengZhu}, which could naturally generate examples of Lie 2-bialgebras.

              The second aim of this paper is to construct solutions of the 2-graded CYBE. We categorify the relation between $\mathcal O$-operators and solutions of the CYBE, and establish the following commutative diagram:
                 $$
\xymatrix{
\mathcal O\mbox{-operators~ on~ Lie~ 2-algebras}\ar[r]& \mbox{solutions~of ~2-graded~CYBE}\ar[r]^{}& \mbox{Lie~2-bialgebras}\\
 \mathcal O\mbox{-operators~ on~ Lie~ algebras} \ar[r]\ar[u]_{\mbox{categorification}}&  \mbox{solutions~of~CYBE}\ar[u]_{\mbox{categorification}}\ar[r]& \mbox{Lie~ bialgebras.}\ar[u]_{\mbox{categorification}}}
                $$
We also find that there are pre-Lie 2-algebras behind the construction of Lie 2-bialgebras in \cite{BaiShengZhu}.

  The paper is organized as follows. In Section 2, we recall Lie 2-algebras and their representations, pre-Lie algebras and their cohomologies, $\mathcal O$-operators and solutions of the  CYBE. In Section 3, first we prove that a 2-term pre-Lie$_\infty$-algebra could give rise to a Lie 2-algebra with a natural representation on itself. Then we introduce the notion of a pre-Lie 2-algebra. At last, we prove that the category of 2-term pre-Lie$_\infty$-algebras and the category of pre-Lie 2-algebras are equivalent (Theorem \ref{thm:equivalent}). In Section 4, we study skeletal pre-Lie 2-algebras and strict pre-Lie 2-algebras in detail. Skeletal pre-Lie 2-algebras could be classified by the third cohomology (Theorem \ref{thm:classify}). We find that there is a natural 3-cocycle associated to a pre-Lie algebra with a skew-symmetric invariant bilinear form. By this fact, we construct a natural example of skeletal pre-Lie 2-algebras associated to a pre-Lie algebra with a skew-symmetric invariant bilinear form. We also introduce the notion of crossed modules of pre-Lie algebras and prove that there is a one-to-one correspondence between crossed modules of pre-Lie algebras and strict pre-Lie 2-algebras (Theorem \ref{thm:one-to-one}). In Section 5, we introduce the notion of an $\mathcal O$-operator on a Lie 2-algebra $\huaG$ associated to a representation $(\huaV;\rho)$, and construct a pre-Lie 2-algebra structure on $\huaV$. In Section 6, we construct solutions of the 2-graded CYBE in the strict Lie 2-algebra $\huaG\ltimes_{\rho^*}\huaV^*$ using $\mathcal O$-operators (Theorem \ref{thm:solution}). In particular, if the strict Lie 2-algebra under consideration is given by a strict pre-Lie 2-algebra,  there is  a natural solution of the $2$-graded CYBE in the strict Lie $2$-algebra $\huaG(\huaA)\ltimes_{(L_0^*,L_1^*)}\huaA^*$ (Theorem \ref{thm:solutionpreLie2}). At last, we give the pre-Lie 2-algebra structure behind the construction of Lie 2-bialgebras in \cite{BaiShengZhu}.

{\bf Acknowledgement:} Y. Sheng gives warmest thanks to Chengming Bai and Chenchang Zhu for very useful comments. Y. Sheng
gratefully acknowledges the support of Courant Research Center
``Higher Order Structures'', G\"{o}ttingen University, where parts of this
work were done during his visit.

\section{Preliminaries}

$\bullet$ Lie 2-algebras and 2-term $L_\infty$-algebras\vspace{2mm}

Vector spaces can be categorified to $2$-vector spaces. A good
introduction for this subject is \cite{baez:2algebras}. Let $\Vect$
be the category of vector spaces.
A {\bf $2$-vector space} is a category in the category $\Vect$.
Thus, a $2$-vector space $C$ is a category with a vector space of
objects $C_0$ and a vector space of morphisms $C_1$, such that all
the structure maps are linear. Let $s,t:C_1\longrightarrow C_0$ be
the source and target maps respectively. Let $\cdot_\ve$ be the
composition of morphisms.

It is well known that the category of 2-vector spaces is
equivalent to the category of 2-term complexes of vector spaces.
Roughly speaking, given a 2-vector space $C$,
$\Ker(s)\stackrel{t}{\longrightarrow}C_0$ is a 2-term complex.
Conversely, any 2-term complex of vector spaces
$\huaV:V_1\stackrel{\dM}{\longrightarrow}V_0$ gives rise to a
2-vector space of which the set of objects is $V_0$, the set of
morphisms is $V_0\oplus V_1$, the source map $s$ and the target map $t$ are given by
$$s(u+m)=u,\quad t(u+m)=u+\dM m,\quad \forall u,v\in V_0,m\in V_1. $$
The composition of morphisms is given by
$$
(u+m)\cdot_\ve(v+n)=(u+m+n),\quad \forall u,v\in V_0, m,n\in V_1,~ \mbox{safisfying} ~v=u+\dM m.
$$
 We denote the 2-vector space associated
to the 2-term complex of vector spaces
$\huaV:V_1\stackrel{\dM}{\longrightarrow}V_0$ by $\V$:
\begin{equation}\label{eqn:V}
\V=\begin{array}{c}
\V_1:=V_0\oplus V_1\\
\vcenter{\rlap{s }}~\Big\downarrow\Big\downarrow\vcenter{\rlap{t }}\\
\V_0:=V_0.
 \end{array}\end{equation}

In this paper, we always assume that a 2-vector space is of the
above form. The identity-assigning map $1:\V_0\longrightarrow\V_1$ is given by $1_u=(u,0)$, for all $u\in \V_0$.

\begin{defi}{\rm \cite{baez:2algebras}}\label{defi:Lie2}
A  Lie $2$-algebra is a $2$-vector space $C$ equipped with
\begin{itemize}
\item[$\bullet$] a skew-symmetric bilinear functor, the bracket, $\Courant{\cdot,\cdot}:C\times C\longrightarrow
C$,
\item[$\bullet$] a skew-symmetric trilinear natural isomorphism, the
Jacobiator,
$$
J_{x,y,z}:\Courant{\Courant{x,y},z}\longrightarrow \Courant{x,\Courant{y,z}}+\Courant{\Courant{x,z},y},
$$
\end{itemize}
such that the following Jacobiator identity is satisfied,
\begin{eqnarray*}
&J_{\Courant{w,x},y,z}\cdot_\ve(\Courant{J_{w,x,z},y}+1)\cdot_\ve(J_{w,\Courant{x,z},y}+J_{\Courant{w,z},x,y}+J_{w,x,\Courant{y,z}})\\
&=\Courant{J_{w,x,y},z}\cdot_\ve(J_{\Courant{w,y},x,z}+J_{w,\Courant{x,y},z})\cdot_\ve(\Courant{J_{w,y,z},x}+1)\cdot_\ve(\Courant{w,J_{x,y,z}}+1).
\end{eqnarray*}
\end{defi}

\begin{defi}
A $2$-term $L_\infty$-algebra structure on a graded vector space $\huaG=\g_0\oplus \g_1$ consists of the following data:
\begin{itemize}
\item[$\bullet$] a linear map $\frkd:\g_1\stackrel{}{\longrightarrow}\g_0,$

\item[$\bullet$] a skew-symmetric bilinear map $\frkl_2:\g_i\times \g_j\longrightarrow
\g_{i+j},~0\leq i+j\leq 1$,

\item[$\bullet$] a  skew-symmetric trilinear map $\frkl_3:\wedge^3 \g_0\longrightarrow
\g_1$,
   \end{itemize}
   such that for any $x_i,x,y,z\in \g_0$ and $m,n\in \g_1$, the following equalities are satisfied:
\begin{itemize}
\item[$\rm(i)$] $\dM \frkl_2(x,m)=\frkl_2(x,\dM m),\quad \frkl_2(\dM m,n)=\frkl_2(m,\dM n),$
\item[$\rm(ii)$]$\dM \frkl_3(x,y,z)=\frkl_2(x,\frkl_2(y,z))+\frkl_2(y,\frkl_2(z,x))+\frkl_2(z,\frkl_2(x,y)),$
\item[$\rm(iii)$]$ \frkl_3(x,y,\dM m)=\frkl_2(x,\frkl_2(y,m))+\frkl_2(y,\frkl_2(m,x))+\frkl_2(m,\frkl_2(x,y)),$
\item[$\rm(iv)$]the Jacobiator identity:\begin{eqnarray*}
&&\sum_{i=1}^4(-1)^{i+1}\frkl_2(x_i,\frkl_3(x_1,\cdots,\widehat{x_i},\cdots,x_4))\\
&&+\sum_{i<j}(-1)^{i+j}\frkl_3(\frkl_2(x_i,x_j),x_1,\cdots,\widehat{x_i},\cdots,\widehat{x_j},\cdots,x_4)=0.\end{eqnarray*}
   \end{itemize}
   \end{defi}
Usually, we denote a $2$-term $L_\infty$-algebra by $(\g_0,\g_1,\frkd,\frkl_2,\frkl_3)$, or simply by $\huaG$. A 2-term $L_\infty$-algebra is called {\bf strict} if $\frkl_3=0.$ Associated to a strict 2-term $L_\infty$-algebra, there is a semidirect product Lie algebra  $\g_0\ltimes \g_1=(\frkg_0\oplus \frkg_{-1},[\cdot,\cdot]_s)$, where the bracket  $[\cdot,\cdot]_s$ is given
by
\begin{equation}\label{eq:semidirect}
[x+m,y+n]_s:=\frkl_2(x,y)+\frkl_2(x,n)+\frkl_2(m,y).
\end{equation}

\begin{defi}\label{defi:Lie-2hom}
Let $\huaG=(\g_0,\g_1,\frkd,\frkl_2,\frkl_3)$ and $\huaG'=(\g_0',\g_1',\frkd',\frkl_2',\frkl_3')$ be $2$-term $L_\infty$-algebras. A
 homomorphism $F$ from $\huaG$ to $ \huaG'$ consists of:
 linear maps $F_0:\g_0\rightarrow \g_0',~F_1:\g_{1}\rightarrow \g_{1}'$
 and $\huaF_{2}:\g_{0}\wedge \g_0\rightarrow \g_{1}'$,
such that the following equalities hold for all $ x,y,z\in \g_{0},
a\in \g_{1},$
\begin{itemize}
\item [$\rm(i)$] $F_0\circ\frkd=\frkd'\circ F_1$,
\item[$\rm(ii)$] $F_{0}\frkl_2(x,y)-\frkl'(F_{0}(x),F_{0}(y))=\frkd'\huaF_{2}(x,y),$
\item[$\rm(iii)$] $F_{1}\frkl_2(x,a)-\frkl'(F_{0}(x),F_{1}(a))=\huaF_{2}(x,\frkd a)$,
\item[$\rm(iv)$]
$\huaF_2(\frkl_2(x,y),z)+c.p.+F_1(\frkl_3(x,y,z))=\frkl_2'(F_0(x),\huaF_2(y,z))+c.p.+\frkl_3'(F_0(x),F_0(y),F_0(z))$.
\end{itemize}\end{defi}

It is well-known that the category of Lie 2-algebras and the category of 2-term $L_\infty$-algebras are equivalent. Thus, when we say ``a Lie 2-algebra'', we mean a $2$-term $L_\infty$-algebra in the sequel.\vspace{2mm}

Let $\huaV:V_1\stackrel{\dM}{\longrightarrow}V_0$ be a complex of
vector spaces. Define $\End^0_\dM(\huaV)$ by
$$
\End^0_\dM(\huaV)\triangleq\{(A_0,A_1)\in\gl(V_0)\oplus
\gl(V_1)|A_0\circ\dM=\dM\circ A_1\},
$$
and define $\End^1(\huaV)\triangleq \Hom(V_0,V_1)$. There is a
differential $\delta:\End^1(\huaV)\longrightarrow \End^0_\dM(\huaV)$
given by
$$
\delta(\phi)\triangleq\phi\circ\dM+\dM\circ\phi,\quad\forall~\phi\in\End^1(\huaV),
$$
and a bracket operation $[\cdot,\cdot]$ given by the graded
commutator. More precisely,  for any $A=(A_0,A_1),B=(B_0,B_1)\in
\End^0_\dM(\huaV)$ and $\phi\in\End^1(\huaV)$, $[\cdot,\cdot]$ is
given by
\begin{eqnarray*}
  [A,B]=A\circ B-B\circ A=(A_0\circ B_0-B_0\circ A_0,A_1\circ B_1-B_1\circ
  A_1),
\end{eqnarray*}
and
\begin{equation}\label{representation}
  ~[A,\phi]=A\circ \phi-\phi\circ A=A_1\circ \phi-\phi\circ A_0.
\end{equation}
These two operations make $\End^1(\huaV)\xrightarrow{\delta}
\End^0_\dM(\huaV)$ into a strict Lie 2-algebra, which we denote by $\End(\huaV)$. It plays
the same role as $\gl(V)$ for a vector space $V$ (\cite{LadaMarkl}).

A {\bf representation} of a Lie 2-algebra $\huaG$ on $\huaV$ is a homomorphism $(\rho_0,\rho_1,\rho_2)$ from $\huaG$ to $\End(\huaV)$. A representation of a strict Lie 2-algebra $\huaG$ on $\huaV$ is called {\bf strict} if $\rho_2=0.$ Given a strict representation of a strict Lie 2-algebra $\huaG$ on $\huaV$, there is a semidirect product strict Lie 2-algebra $\huaG\ltimes\huaV$, in which the degree 0 part is $\g_0\oplus V_0,$ the degree 1 part is $\g_1\oplus V_1$, the differential is $\frkd+\dM:\g_1\oplus V_1\longrightarrow \g_0\oplus V_0$, and for all $x,y\in\g_0, a\in\g_1, u,v\in V_0, m\in V_1,$ $\frkl_2^s$ is given by
\begin{eqnarray*}
\frkl_2^s(x+u,y+v)&=&\frkl_2(x,y)+\rho_0(x)v-\rho_0(y)u,\\
\frkl_2^s(x+u,a+m)&=&\frkl_2(x,a)+\rho_0(x)m-\rho_1(a)u.
\end{eqnarray*}

$\bullet$ Pre-Lie algebras and their representations

\begin{defi}  A {\bf pre-Lie algebra} $(A,\cdot_A)$ is a vector space $A$ equipped with a bilinear product $\cdot_A:\otimes^2A\longrightarrow A$
such that for any $x,y,z\in A$, the associator
$(x,y,z)=(x\cdot_A y)\cdot_A-x\cdot_A(y\cdot_A z)$ is symmetric in $x,y$,
i.e.,
$$(x,y,z)=(y,x,z),\;\;{\rm or}\;\;{\rm
equivalently,}\;\;(x\cdot_A y)\cdot_A z-x\cdot_A(y\cdot_A z)=(y\cdot_A x)\cdot_A
z-y\cdot_A(x\cdot_A z).$$
\end{defi}

 Let $A$ be a pre-Lie algebra. The commutator
$ [x,y]_A=x\cdot_A y-y\cdot_A x$ defines a Lie algebra structure on $A$,
which is called the {\bf sub-adjacent Lie algebra} of $A$ and denoted by $\g(A)$. Furthermore, $L:A\rightarrow
\gl(A)$ with $L_xy=x\cdot_A y$ gives a representation of the Lie
algebra $\g(A)$ on $A$. See \cite{leftsymm4} for more details.

\begin{defi}
Let $(A,\cdot_A)$ be a pre-Lie algebra and $V$  a vector
space. A {\bf representation} of $A$ on $V$ consists of a pair
$(\rho,\mu)$, where $\rho:A\longrightarrow \gl(V)$ is a representation
of the Lie algebra $\g(A)$ on $V $ and $\mu:A\longrightarrow \gl(V)$ is a linear
map satisfying \begin{eqnarray}\label{representation condition 2}
 \rho(x)\mu(y)u-\mu(y)\rho(x)u=\mu(x\cdot_A y)u-\mu(y)\mu(x)u, \quad \forall x,y\in A,~ u\in V.
\end{eqnarray}
\end{defi}
Usually, we denote a representation by $(V;\rho,\mu)$. In this case, we will also say that $(\rho,\mu)$ is an {\bf action} of $(A,\cdot_A)$ on $V$. Define $R:A\longrightarrow\gl(A)$ by $R_xy=y\cdot_Ax$. Then $(A;L,R)$ is a representation of $(A,\cdot_A)$. Furthermore, $(A^*;\ad^*=L^*-R^*,-R^*)$ is also a representation of $(A,\cdot_A)$, where $L^*$ and $R^*$ are given by
$$
\langle L^*_x\xi,y\rangle=\langle\xi,-L_xy\rangle,\quad \langle R^*_x\xi,y\rangle=\langle\xi,-R_xy\rangle,\quad \forall x,y\in A,\xi\in A^*.
$$

The cohomology complex for a pre-Lie algebra $(A,\cdot_A)$ with a representation $(V;\rho,\mu)$ is given as follows (\cite{cohomology of pre-Lie}).
The set of $(n+1)$-cochains is given by
$$C^{n+1}(A,V)=\Hom(\wedge^{n}A\otimes A,V),\
n\geq 0.$$  For all $\omega\in C^{n}(A,E)$, the coboundary operator $d:C^{n}(A,E)\longrightarrow C^{n+1}(A,E)$ is given by
 \begin{eqnarray*}
 &&d\omega(x_1,x_2,\cdots,x_{n+1})\\
 &=&\sum_{i=1}^{n}(-1)^{i+1}\rho(x_i)\omega(x_1,x_2,\cdots,\hat{x_i},\cdots,x_{n+1})\\
 &&+\sum_{i=1}^{n}(-1)^{i+1}\mu(x_{n+1})\omega(x_1,x_2,\cdots,\hat{x_i},\cdots,x_n,x_i)\\
 &&-\sum_{i=1}^{n}(-1)^{i+1}\omega(x_1,x_2,\cdots,\hat{x_i},\cdots,x_n,x_i\cdot_A x_{n+1})\\
 &&+\sum_{1\leq i<j\leq n}(-1)^{i+j}\omega([x_i,x_j]_A,x_1,\cdots,\hat{x_i},\cdots,\hat{x_j},\cdots,x_{n+1}),
\end{eqnarray*}
for all $x_i\in \Gamma(A),i=1,2\cdots,n+1$.\vspace{3mm}

$\bullet$ $\mathcal O$-operators and solutions of the Classical Yang-Baxter Equations\vspace{2mm}

Let $(\frak g,[\cdot,\cdot]_\g)$ be a Lie algebra and $( V;\rho)$ be a representation. A linear map $T:V\rightarrow \g$ is called an {\bf $\mathcal O$-operator} on $\g$ associated to the representation  $( V;\rho)$ if $T$ satisfies
\begin{equation}
  [T(u),T(v)]_\g=T(\rho (T(u))v-\rho(T(v))u),\quad \forall~ u,v\in V.
\end{equation}

Associated to a representation $(V;\rho)$, we
 have the semidirect product Lie algebra $\g\ltimes_{\rho^*}V^*$, where $\rho^*:\g\longrightarrow\gl(V^*)$ is the dual representation. A linear map $T:V\longrightarrow\g$ can be view as an element $\overline{T}\in\otimes^2(\g\oplus V^*)$ via
\begin{equation}\label{eq:Tbar}
\overline{T}(\xi+u,\eta+v)=\langle T(u),\eta\rangle,\quad \forall~\xi+u,\eta+v\in\g^*\oplus V.
\end{equation}
Let $\sigma$ be the exchange operator acting on
 the tensor space, then $r\triangleq\overline{T}-\sigma (\overline{T})$ is skew-symmetric.

\begin{thm}\label{thm:O-operator}
Let $T:V\rightarrow \frak g$ be a linear map. Then
 $r=\overline{T}-\sigma (\overline{T})$ is a  solution of the classical Yang-Baxter equation
in the Lie algebra $\frak g\ltimes_{\rho^*} V^*$ if and only if $T$ is an $\mathcal O$-operator.
\end{thm}
\begin{thm}{\rm \cite{Bai:CYBE}}
Let $A$ be a left-symmetric algebra. Then
\begin{equation}
r=\sum_{i=1}^n (e_i\otimes e_i^*-e_i^*\otimes
e_i)\label{eq:rrr}\end{equation} is a solution of the CYBE in $\frak
g(A) \ltimes_{L^*} A^*$, where $\{e_i\}$ is a basis of $A$, and $\{e_i^*\}$ is the dual basis.
\end{thm}

\section{ 2-term pre-Lie$_\infty$-algebras and Pre-Lie 2-algebras}
In this section, we show that a 2-term pre-Lie$_\infty$-algebra could give rise to a Lie 2-algebra, and the left multiplication gives rise to a representation of the Lie 2-algebra. We introduce the notion of a pre-Lie 2-algebra, which is a categorification of a pre-Lie algebra. We prove that the category of 2-term pre-Lie$_\infty$-algebras and the category of pre-Lie 2-algebras are equivalent.

\subsection{2-term Pre-Lie$_\infty$-algebras}
The terminology of a pre-Lie$_\infty$-algebra is introduced in \cite{chapoton}, which is a  right-symmetric algebra up to homotopy. By a slight modification, we could obtain a  left-symmetric algebra (the pre-Lie algebra we use in this paper) up to homotopy. By truncation, we obtain a $2$-term pre-Lie$_\infty$-algebra.

\begin{defi}\label{defi:2-term pre}
  A {\bf $2$-term pre-Lie$_\infty$-algebra} is a $2$-term graded vector spaces $\huaA=A_0\oplus A_1$, together with linear maps $\dM:A_1\longrightarrow A_0$, $\cdot:A_i\otimes A_j\longrightarrow A_{i+j}$, $0\leq i+j\leq 1$, and $l_3:\wedge^2A_0\otimes A_0\longrightarrow A_1$, such that for all $v,v_i\in A_0$ and $m,n\in A_1$, we have
  \begin{itemize}
    \item[\rm($a_1$)] $\dM (v\cdot m)=v\cdot \dM m,$
\item[\rm($a_2$)] $ \dM(m\cdot v)=(\dM m)\cdot v,$
\item[\rm($a_3$)]$ \dM m\cdot n=m\cdot \dM n,$
    \item[\rm($b_1$)]$v_0\cdot(v_1\cdot v_2)-(v_0\cdot v_1)\cdot v_2-v_1\cdot(v_0\cdot v_2)+(v_1\cdot v_0)\cdot v_2=\dM l_3(v_0,v_1,v_2),$

     \item[\rm($b_2$)]$v_0\cdot(v_1\cdot m)-(v_0\cdot v_1)\cdot m-v_1\cdot(v_0\cdot m)+(v_1\cdot v_0)\cdot m=l_3(v_0,v_1,\dM m),$

     \item[\rm($b_3$)]$m\cdot(v_1\cdot v_2)-(m\cdot v_1)\cdot v_2-v_1\cdot(m\cdot v_2)+(v_1\cdot m)\cdot v_2=l_3(\dM  m,v_1,v_2),$

    \item[\rm(c)]\begin{eqnarray*}
    &&v_0\cdot l_3(v_1,v_2,v_3)- v_1\cdot l_3(v_0,v_2,v_3)+ v_2\cdot l_3(v_0,v_1,v_3)\\
    &&+l_3(v_1,v_2,v_0)\cdot v_3- l_3(v_0,v_2,v_1)\cdot v_3+ l_3(v_0,v_1,v_2)\cdot v_3\\&&-l_3( v_1,v_2,v_0\cdot v_3)+l_3( v_0,v_2,v_1\cdot v_3)-l_3(v_0,v_1, v_2\cdot v_3)\\
    &&-l_3(v_0\cdot v_1-v_1\cdot  v_0, v_2,v_3)+l_3(v_0\cdot v_2-v_2\cdot v_0,v_1,v_3)-l_3(v_1\cdot v_2-v_2\cdot v_1,v_0,v_3)=0.
    \end{eqnarray*}
  \end{itemize}
\end{defi}
Usually, we denote a 2-term pre-Lie$_\infty$-algebra by $(A_0,A_1,\dM,\cdot,l_3)$, or simply by $\huaA$. A 2-term pre-Lie$_\infty$-algebra $(A_0,A_1,\dM,\cdot,l_3)$ is said to be {\bf skeletal } ({\bf strict}) if $\dM=0$ $(l_3=0)$.

Given a 2-term pre-Lie$_\infty$-algebra  $(A_0,A_1,\dM,\cdot,l_3)$, we
 define $\frkl_2:A_i\wedge A_j\longrightarrow A_{i+j}$ and $\frkl_3:\wedge^3A_0\longrightarrow A_1$ by
\begin{eqnarray}
  \label{eq:l21}\frkl_2(u,v)&=&u\cdot v-v\cdot u,\\
   \label{eq:l22} \frkl_2(u,m)&=&-\frkl_2(m,u)=u\cdot m-m\cdot u,\\
   \label{eq:l3}\frkl_3(u,v,w)&=&l_3(u,v,w)+l_3(v,w,u)+l_3(w,u,v).
\end{eqnarray}
Furthermore, define $L_0:A_0\longrightarrow\End(A_0)\oplus \End(A_1)$ by
\begin{equation}\label{eq:L0}
 L_0(u)v=u\cdot v,\quad L_0(u)m=u\cdot m.
\end{equation}
Define $L_1:A_1\longrightarrow\Hom(A_0,A_1)$ by
\begin{equation}\label{eq:L1}
 L_1(m)u=m\cdot u.
\end{equation}
Define $L_2:\wedge^2A_0\longrightarrow \Hom(A_0,A_1)$ by
\begin{equation}\label{eq:L2}
 L_2(u,v)w=-l_3(u,v,w),\quad \forall u,v,w\in A_0.
\end{equation}
\begin{thm}\label{thm:main1}
 Let $\huaA=(A_0,A_1,\dM,\cdot,l_3)$ be a $2$-term pre-Lie$_\infty$-algebra. Then, $(A_0,A_1,\dM,\frkl_2,\frkl_3)$ is a Lie $2$-algebra, which we denote by $\huaG(\huaA)$, where $\frkl_2$ and $\frkl_3$ are given by \eqref{eq:l21}-\eqref{eq:l3} respectively. Furthermore, $(L_0,L_1,L_2)$ is a representation of the Lie $2$-algebra $\huaG(\huaA)$ on the complex of vector spaces $A_1\stackrel{\dM}{\longrightarrow} A_0$, where $L_0,L_1,L_2$ are given by \eqref{eq:L0}-\eqref{eq:L2} respectively.
\end{thm}
\pf By Conditions ($a_1$)-($a_3$), we have
\begin{eqnarray*}
  \dM \frkl_2(v,m)&=&\dM(v\cdot m-m\cdot v)=v\cdot\dM m-(\dM m)\cdot v=\frkl_2(v,\dM m),\\
  \frkl_2(\dM m, n)&=&(\dM m)\cdot n-n\cdot \dM m=m\cdot\dM n-(\dM n)\cdot m=\frkl_2(m,\dM n).
\end{eqnarray*}
By Condition ($b_1$), we have
\begin{eqnarray*}
\frkl_2(v_0,\frkl_2(v_1,v_2))+c.p.&=&\frkl_2(v_0,v_1\cdot v_2-v_2\cdot v_1)+c.p.\\
  &=&v_0\cdot(v_1\cdot v_2)-(v_1\cdot v_2)\cdot v_0-v_0\cdot(v_2\cdot v_1)+(v_2\cdot v_1)\cdot v_0+c.p.\\
  &=&\dM (l_3(v_0,v_1,v_2)+l_3(v_1,v_2,v_0)+l_3(v_2,v_0,v_1))\\
  &=&\dM \frkl_3(v_0,v_1,v_2).
  \end{eqnarray*}
Similarly, by Conditions ($b_2$) and  ($b_3$), we have
  \begin{eqnarray*}
&&\frkl_2(v_0,\frkl_2(v_1,m))+\frkl_2(v_1,\frkl_2(m,v_0))+\frkl_2(m,\frkl_2(v_0,v_1))\\&=&\frkl_2(v_0,v_1\cdot m-m\cdot v_1)+\frkl_2(v_1,m\cdot v_0-v_0\cdot m)+\frkl_2(m,v_0\cdot v_1-v_1\cdot v_0)\\
  &=&v_0\cdot(v_1\cdot m)-(v_1\cdot m)\cdot v_0-v_0\cdot(m\cdot v_1)+(m\cdot v_1)\cdot v_0\\
  &&v_1\cdot(m\cdot v_0)-(m\cdot v_0)\cdot v_1-v_1\cdot(v_0\cdot m)+(v_0\cdot m)\cdot v_1\\
  && m\cdot(v_0\cdot v_1)-(v_0\cdot v_1)\cdot m-m\cdot(v_1\cdot v_0)+(v_1\cdot v_0)\cdot m     \\
  &=&\dM (l_3(v_0,v_1,\dM m)+l_3(v_1,\dM m,v_0)+l_3(\dM m,v_0,v_1))\\
  &=& \frkl_3(v_0,v_1,\dM m).
\end{eqnarray*}
At last, by Condition (c), we could get
\begin{eqnarray*}
  &&\frkl_2(v_0,\frkl_3(v_1,v_2,v_3))-\frkl_2(v_1,\frkl_3(v_0,v_2,v_3))+\frkl_2(v_2,\frkl_3(v_0,v_1,v_3))-\frkl_2(v_3,\frkl_3(v_0,v_1,v_2))\\
  &=&\frkl_3(\frkl_2(v_0,v_1),v_2,v_3)-\frkl_3(\frkl_2(v_0,v_2),v_1,v_3)+\frkl_3(\frkl_2(v_0,v_3),v_1,v_2)+\frkl_3(\frkl_2(v_1,v_2),v_0,v_3)
  \\&&-\frkl_3(\frkl_2(v_1,v_3),v_0,v_2)+\frkl_3(\frkl_2(v_2,v_3),v_0,v_1).
\end{eqnarray*}
Thus, $(A_0,A_1,\dM,\frkl_2,\frkl_3)$ is a Lie $2$-algebra.

By Condition ($a_1$), we deduce that $L_0(u)\in \End^0_\dM(\huaA)$ for all $u\in A_0$. By Conditions ($a_2$) and ($a_3$), we have
\begin{equation}\label{eq:conmor1}
\delta\circ L_1(m)=L_0(\dM m).
\end{equation}
Furthermore, we have
\begin{eqnarray*}
  L_0({\frkl_2(u,v)})w&=&(u\cdot v)\cdot w-(v\cdot u)\cdot w=u\cdot(v\cdot w)-v\cdot(u\cdot w)-\dM l_3(u,v,w)\\
  &=&[L_0(u),L_0(v)]w-\dM l_3(u,v,w),
\end{eqnarray*}
which implies that
\begin{equation}\label{eq:conmor2}
  L_0({\frkl_2(u,v)})-[L_0(u),L_0(v)]=\dM\circ L_2(u,v).
\end{equation}
Similarly, we have
\begin{equation}\label{eq:conmor3}
  L_1({\frkl_2(u,m)})-[L_0(u),L_1(m)]= L_2(u,\dM  m).
\end{equation}
At last, by Condition (c) in Definition \ref{defi:2-term pre}, we get
\begin{equation}\label{eq:conmor4}
 ~- [L_0(u),L_2(v,w)]+L_2(\frkl_2(u,v),w)+c.p.+L_1({\frkl_3(u,v,w)})=0.
\end{equation}
By \eqref{eq:conmor1}-\eqref{eq:conmor4}, we deduce that $(L_0,L_1,L_2)$ is a homomorphism from the Lie $2$-algebra $\huaG(\huaA)$ to $\End(\huaV)$. The proof is finished.\qed

\begin{defi}
 Let    $\huaA=(A_0,A_1,\dM,\cdot,l_3)$ and $\huaA'=(A_0',A_1',\dM',\cdot',l_3')$ be $2$-term pre-Lie$_\infty$-algebras. A homomorphism $(F_0,F_1,F_2)$ from $\huaA$ to $\huaA'$ consists of linear maps $F_0:A_0\longrightarrow A_0'$,  $F_1:A_1\longrightarrow A_1'$, and $F_2:A_0\otimes A_0\longrightarrow A_1'$ such that the following equalities hold:
 \begin{itemize}
  \item[\rm(i)] $F_0\circ \dM=\dM'\circ F_1,$
   \item[\rm(ii)] $F_0(u\cdot v)-F_0(u)\cdot'F_0( v)=\dM'F_2(u,v),$
    \item[\rm(iii)] $F_1(u\cdot m)-F_0(u)\cdot'F_1( m)=F_2(u,\dM m),\quad F_1( m\cdot u)-F_1( m)\cdot'F_0(u)=F_2(\dM m,u),$
     \item[\rm(iv)] $F_0(u)\cdot'F_2(v,w)-F_0(v)\cdot'F_2(u,w)+F_2(v,u)\cdot'F_0(w)-F_2(u,v)\cdot'F_0(w)-F_2(v,u\cdot w)\\+F_2(u,v\cdot w)-F_2(u\cdot v ,w)+F_2(v\cdot u ,w)+l_3'(F_0(u),F_0(v),F_0(v))-F_1l_3(u,v,w)=0$.
 \end{itemize}
\end{defi}
By straightforward computations, we have
\begin{pro}
  Let $\huaA=(A_0,A_1,\dM,\cdot,l_3)$ to $\huaA'=(A_0',A_1',\dM',\cdot',l_3')$ be $2$-term pre-Lie$_\infty$-algebras, and $(F_0,F_1,F_2)$ be a  homomorphism from $\huaA$ to $\huaA'$. Then, $(F_0,F_1,\huaF_2)$ is a  homomorphism from the corresponding Lie $2$-algebra $\huaG(\huaA)$ to $\huaG(\huaA')$, where $\huaF_2:\wedge^2 A_0\longrightarrow A_1$ is given by
  \begin{equation}
    \huaF_2(u,v)=F_2(u,v)-F_2(v,u).
  \end{equation}
\end{pro}

At the end of this section, we introduce composition and identity for $2$-term pre-Lie$_\infty$-algebra homomorphisms. Let $F=(F_0,F_1,F_2):\huaA\longrightarrow\huaA'$ and $G=(G_0,G_1,G_2):\huaA'\longrightarrow\huaA''$ be $2$-term pre-Lie$_\infty$-algebra homomorphisms. Their composition $GF=((GF)_0,(GF)_1,(GF)_2)$ is defined by $(GF)_0=G_0\circ F_0,(GF)_1=G_1\circ F_1,$  and $(GF)_2$ is given by
\begin{equation}
  (GF)_2(u,v)=G_2(F_0(u),F_0(v))+G_1(F_2(u,v)).
\end{equation}
It is straightforward to verify that $GF=((GF)_0,(GF)_1,(GF)_2):\huaA\longrightarrow\huaA''$ is a $2$-term pre-Lie$_\infty$-algebra homomorphism. It is obvious that $(\id_{A_0},\id_{A_1},0)$ is the identity homomorphism. Thus, we obtain
\begin{pro}\label{pro:2preLie}
  There is a category, which we denote by ${\rm\bf 2preLie}$, with $2$-term pre-Lie$_\infty$-algebras as objects, homomorphisms between them as morphisms.
\end{pro}

\subsection{Pre-Lie 2-algebras}

\begin{defi}
  A pre-Lie  $2$-algebra is a $2$-vector space $\V$ endowed with
   a bilinear functor
  $\star:\V\times\V\longrightarrow\V$ and a natural isomorphism $J_{u,v,w}$ for
all $u,v,w\in\V_0$,
\begin{equation}\label{defiLeibniz21}
J_{u,v,w}:(u\star v)\star w-u\star(v\star w)\longrightarrow (v\star u)\star w-v\star(u\star w),
\end{equation}
such that the following  identity is satisfied:{\footnotesize
\begin{eqnarray}
\nonumber&&(0J_{1,2,3})\cdot_\ve(J_{0,21,3}+1_{(0(21))3}+J_{0,2,13}-1_{(02)(13)})\\
\nonumber&&\cdot_\ve(1_{(0(21))3-((21)0)3+(21)(03)}+J_{02,1,3}-1_{((02)1)3}-J_{20,1,3}+1_{((20)1)3}+2J_{0,1,3}-21_{(01)3})\\
\nonumber&=&(-J_{0,12,3}+1_{(0(12))3}+J_{0,1,23}-1_{(01)(23)})\\
\nonumber&&\cdot_\ve(J_{1,2,03}+1_{1(2(03))}+J_{01,2,3}-1_{((01)2)3}-J_{10,2,3}+1_{((10)2)3}+1J_{0,2,3}-11_{(02)3}+1_{-((12)0)3+(0(12))3})\\
\nonumber&&\cdot_\ve(-J_{1,2,0}3-1_{1(20)+2(10)}3-J_{2,0,1}3+1_{(20)1}3-J_{0,1,2}3+1_{(10)2}3\\
&&+1_{(21)(03)-2(1(03))-2((01)3)+2((10)3)-1((02)3)+1((20)3)}).
\end{eqnarray}}
Here, $0,1,2,3$ denote $v_0,v_1,v_2,v_3$ respectively, $ij$ denotes $v_i\star v_j$, $iJ_{j,k,l}$ denotes $v_i\star J_{v_j,v_k,v_l}$, and $J_{j,k,l}i$ denotes $ J_{v_j,v_k,v_l}\star v_i$.
Or, in terms of a commutative diagram,{\footnotesize
$$
\xymatrix{0((12)3-1(23))\ar[d]_{-J_{0,12,3}+1_{(0(12))3}+J_{0,1,23}-1_{(01)(23)}}\ar[rr]^{0J_{1,2,3}}&&0((21)3-2(13))\ar[d]^{J_{0,21,3}+1_{(0(21))3}+J_{0,2,13}-1_{(02)(13)}}\\
P\ar[d]_{\epsilon}&&Q\ar[d]^{\varepsilon}\\
M\ar[rr]^{\kappa}&&N}
$$
where
\begin{eqnarray*}
  P&=&-((12)0)3+(12)(03)+(0(12))3-(01)(23)+(10)(23)-1(0(23)),\\
  Q&=&(0(21))3-((21)0)3+(21)(03)-(02)(13)+(20)(13)-2(0(13)),\\
  \epsilon&=&J_{1,2,03}+1_{1(2(03))}+J_{01,2,3}-1_{((01)2)3}-J_{10,2,3}+1_{((10)2)3}+1J_{0,2,3}-11_{(02)3}+1_{-((12)0)3+(0(12))3},\\
  \varepsilon&=&1_{(0(21))3-((21)0)3+(21)(03)}+J_{02,1,3}-1_{((02)1)3}-J_{20,1,3}+1_{((20)1)3}+2J_{0,1,3}-21_{(01)3},\\
  \kappa&=&-J_{1,2,0}3-1_{1(20)+2(10)}3-J_{2,0,1}3+1_{(20)1}3-J_{0,1,2}3+1_{(10)2}3\\
  &&+1_{(21)(03)-2(1(03))-2((01)3)+2((10)3)-1((02)3)+1((20)3)},\\
  M&=&-((12)0)3+(0(12))3+1(2(03))+(21)(03)-2(1(03))-((01)2)3+(2(01))3-2((01)3)\\
  &&+((10)2)3-(2(10))3+2((10)3)-1((02)3)+1((20)3)-1(2(03)),\\
  N&=& (0(21))3-((21)0)3+(21)(03)-((02)1)3+(1(02))3-1((02)3)\\
  &&+((20)1)3-(1(20))3+1((20)3)-2((01)3)+2((10)3)-2(1(03)).
\end{eqnarray*}}
\end{defi}

\begin{defi}
Let $(\V,\star,J)$ and  $(\V',\star',J')$ be pre-Lie $2$-algebras. A homomorphism $\Phi:\V\longrightarrow\V'$ consists of
\begin{itemize}
  \item A linear functor $(\Phi_0,\Phi_1)$ from $\V$ to $\V'$,
  \item A bilinear natural transformation $\Phi_2:\Phi_0(u)\star'\Phi_0(v)\longrightarrow\Phi_0(u\star v)$,
   \end{itemize}
   such that the following identity holds:
   \begin{eqnarray*}
     &&J'_{\Phi_0(u),\Phi_0(v),\Phi_0(w)}\cdot_\ve(F_2(v,u)\star 1_{\Phi_0(w)}-1_{\Phi_0(v)}\star F_2(u,w))\cdot_\ve(F_2(v\star u,w)-F_2(v,u\star w))\\
     &=&(F_2(u,v)\star 1_{\Phi_0(w)}-1_{\Phi_0(u)}\star F_2(v,w))\cdot_\ve(F_2(u\star v,w)-F_2(u,v\star w))\cdot_\ve F_1J_{u,v,w},
   \end{eqnarray*}
or, in terms of  a commutative   diagram:{\footnotesize
   $$
\xymatrix{(\Phi_0(u)\star'\Phi_0(v))\star'\Phi_0(w)-\Phi_0(u)\star'(\Phi_0(v)\star'\Phi_0(w))\ar[d]_{F_2(u,v)\star 1_{\Phi_0(w)}-1_{\Phi_0(u)}\star F_2(v,w)}\ar[rr]^{J'_{\Phi_0(u),\Phi_0(v),\Phi_0(w)}}&&(\Phi_0(v)\star'\Phi_0(u))\star'\Phi_0(w)-\Phi_0(v)\star'(\Phi_0(u)\star'\Phi_0(w))\ar[d]^{F_2(v,u)\star 1_{\Phi_0(w)}-1_{\Phi_0(v)}\star F_2(u,w)}\\
\Phi_0(u\star v)\star \Phi_0(w)-\Phi_0(u)\star\Phi_0(v\star w)\ar[d]_{F_2(u\star v,w)-F_2(u,v\star w)}&&\Phi_0(v\star u)\star \Phi_0(w)-\Phi_0(v)\star\Phi_0(u\star w)\ar[d]^{F_2(v\star u,w)-F_2(v,u\star w)}\\
F_0((u\star v)\star w)-F_0(u\star( v\star w))\ar[rr]^{F_1J_{u,v,w}}&&F_0((v\star u)\star w)-F_0(v\star (u\star w)).}
$$}
\end{defi}

The composition of two homomorphisms $\Phi:\V\longrightarrow\V'$ and $\Psi:\V'\longrightarrow\V''$, which we denote by $\Psi\Phi:\V\longrightarrow\V''$ is defined as follows:
$$
(\Psi\Phi)_0=\Psi_0\circ\Phi_0,\quad (\Psi\Phi)_1
=\Psi_1\circ \Phi_1,\quad (\Psi\Phi)_2(u,v)=\Psi_2(\Phi_0(u),\Phi_0(v))\cdot_\ve \Psi_1(\Phi_2(u,v)).$$
The identity homomorphism $1_\V$ has the identity functor as its underlying functor, together with an identity natural transformation. It is straightforward to obtain
\begin{pro}\label{pro:preLie2}
There is a category, which we denote by ${\rm\bf preLie2}$, with pre-Lie $2$-algebras as objects, homomorphisms between them as morphisms.
\end{pro}
\subsection{The equivalence}

\begin{thm}\label{thm:equivalent}
  The categories ${\rm\bf 2preLie}$ and ${\rm\bf preLie2}$, which are given in Proposition \ref{pro:2preLie} and \ref{pro:preLie2} respectively,  are equivalent.
\end{thm}
Thus, in the following sections, a 2-term pre-Lie$_\infty$-algebra will be called a pre-Lie 2-algebra.\vspace{1mm}

\pf  We only give a sketch of the proof. First we construct a functor
$T:{\bf 2preLie}\longrightarrow {\bf preLie2}$.

Given a 2-term
pre-Lie$_\infty$-algebra
$\huaA=(A_0, A_1,\dM,\cdot,l_3)$,
we have a 2-vector space $\mathbb A$ given by \eqref{eqn:V}. More precisely, we have $\A_0=A_0,~\A_1=A_0\oplus A_1.$ Define a bilinear functor $\star:\A\times
\A\longrightarrow \A$ by
$$
(u+m)\star(v+n)=u\cdot v+u\cdot n+ m\cdot v+\dM
m\cdot n,\quad\forall~u+m,v+n\in \A_1=A_0\oplus A_1.
$$
Define the Jacobiator $J:\otimes^3\A_0\longrightarrow \A_1$ by
$$
J_{u,v,w}=(u\cdot v)\cdot w-u\cdot(v\cdot w)+l_3(x,y,z).
$$
 By the various conditions of
$\huaA$
being a 2-term pre-Lie$_\infty$-algebra, we deduce that
$(\A,\star,J)$ is a pre-Lie 2-algebra. Thus, we have
constructed a pre-Lie 2-algebra $\A=T(\huaA)$ from a 2-term
pre-Lie$_\infty$-algebra $\huaA$.

For any homomorphism $F=(F_0,F_1,F_2)$ form $\huaA$ to
$\huaA'$, next we construct a pre-Lie 2-algebra homomorphism $\Phi=T(F)$
from $\A=T(\huaA)$ to $\A'=T(\huaA')$.
Let $\Phi_0=F_0,~\Phi_1=F_0\oplus F_1$, and $\Phi_2$ be given by
$$
\Phi_2(u,v)=F_0(u)\cdot' F_0(v)+F_2(u,v).
$$
Then $\Phi_2(u,v)$ is a natural isomorphism
from $\Phi_0(u)\cdot'\Phi_0(v)$ to $\Phi_0(u\cdot v)$, and $\Phi=(\Phi_0,\Phi_1,\Phi_2)$ is a
homomorphism from $\A$ to $\A'$.

One can also deduce that $T$ preserves the identity homomorphisms and
the composition of homomorphisms. Thus, $T$ constructed above is a
functor from {\bf 2pre-Lie} to {\bf preLie2}.

Conversely, given a pre-Lie 2-algebra $\A$, we construct the 2-term
pre-Lie$_\infty$-algebra $\huaA=S(\A)$ as follows. As a complex of vector
spaces, $\huaA$ is obtained as follows:
$A_0=\A_0,~A_1=\Ker(s)$, and $\dM=t|_{\Ker(s)}$, where $s,t$ are the sauce map and the target map in the 2-vector space $\A$. Define a multiplication $\cdot:A_i\otimes A_j\longrightarrow A_{i+j},~0\leq i+j\leq 1$, by
$$
u\cdot v=u\star v,\quad u\cdot m=1_u\star m,\quad m\cdot u= m \star 1_u,\quad \forall u,v\in A_0,~m,n\in A_1.
$$
Define $l_3:\wedge^2A_0\otimes A_0\longrightarrow A_1$ by
$$
l_3(u,v,w)=J_{u,v,w}-1_{s(J_{u,v,w})}.
$$
 The
various conditions of $\A$ being a pre-Lie 2-algebra imply that
$\huaA$ is 2-term pre-Lie$_\infty$-algebra.

Let $\Phi=(\Phi_0,\Phi_1,\Phi_2):\A\longrightarrow \A'$ be a pre-Lie 2-algebra
homomorphism, and $S(\A)=\huaA,~S(\A')=\huaA'$. Define
$S(\Phi)=F=(F_0,F_1,F_2)$ as follows. Let $F_0=\Phi_0$,
$F_1=\Phi_1|_{A_1=\Ker(s)}$ and define $F_2$ by
$$
F_2(u,v)=\Phi_2(u,v)-1_{s(\Phi_2(u,v))}.
$$
It is not hard to deduce that $F$ is a homomorphism between 2-term pre-Lie$_\infty$-algebras.
Furthermore, $S$ also preserves the identity homomorphisms and the
composition of homomorphisms. Thus, $S$ is a functor from {\bf preLie2} to
{\bf 2pre-Lie}.

We are left to show that there are natural isomorphisms
$\alpha:T\circ S\Longrightarrow \id_{{\bf preLie2}}$ and $\beta:S\circ
T\Longrightarrow \id_{{\bf 2preLie}}$. For a pre-Lie 2-algebra
$(\A,\star,J)$, applying the functor $S$ to $\A$, we
obtain a 2-term pre-Lie$_\infty$-algebra
$\huaA=(A_0,A_1,\dM=t|_{\Ker(s)},\cdot,l_3)$, where $A_0=\A_0,~A_1=\Ker(s)$.
Applying the functor  $T$ to $\huaA$, we obtain a pre-Lie 2-algebra
$(\A',\star',J')$, with the space $A_0$ of objects and
the space $A_0\oplus \Ker(s)$ of morphisms. Define
$\alpha_\A:\A'\longrightarrow \A$ by setting
$$
(\alpha_\A)_0(u)=u,\quad (\alpha_\A)_1(u+m)=1_u+m.
$$
It is obvious that $\alpha_\A$ is an isomorphism of 2-vector spaces.
Furthermore, since $\star$ is a bilinear functor, we have
$1_u\star1_v=1_{u\star v}$, and
$$
m\star n=(m\cdot_\ve 1_{\dM m})\star(1_0\cdot_\ve n)=(m\star1_0)\cdot_\ve(1_{\dM
m}\star n)=1_{\dM m}\star n.
$$
Therefore, we have
\begin{eqnarray*}
  \alpha_\A((u+m)\star'(v+n))&=&\alpha_\A(u\cdot v+u\cdot n+m\cdot v+\dM
  m\cdot n)\\
  &=&\alpha_\A(u\star v+1_u\star n+m\star 1_v+1_{\dM m}\star n)\\
  &=&1_{u\star v}+1_u\star n+m\star 1_v+1_{\dM m}\star n\\
  &=&1_{u}\star 1_{v}+1_u\star n+m\star 1_v+1_{\dM m}\star n\\
    &=&\alpha_\A(u+m)\star\alpha_\A(v+n),
\end{eqnarray*}
which implies that $\alpha_\A$ is also a pre-Lie 2-algebra homomorphism
with $(\alpha_\A)_2$ the identity isomorphism. Thus, $\alpha_\A$ is an
isomorphism of pre-Lie 2-algebras. It is also easy to see that it is
a natural isomorphism.

For a 2-term pre-Lie$_\infty$-algebra
$\huaA=(A_0,A_1,\dM,\cdot,l_3)$,
applying the functor $S$ to $\huaA$, we obtain a pre-Lie 2-algebra
$(\A,\star,J)$. Applying the functor $T$ to $\A$, we obtain
exactly the same 2-term pre-Lie$_\infty$-algebra $\huaA$. Thus,
$\beta_\huaA=\id_\huaA=(\id_{A_0},\id_{A_1})$ is the natural
isomorphism from $T\circ S$ to $\id_{{\bf 2preLie}}$. This finishes
the proof. \qed

\begin{rmk}
  We can further obtain $2$-categories { $\bf 2preLie$} and  {\bf preLie2}  by
  introducing $2$-morphisms and strengthen Theorem
  \ref{thm:equivalent} to the $2$-equivalence of $2$-categories. We omit details.
\end{rmk}

\section{Skeletal and strict  pre-Lie 2-algebras}
In this section, we study  skeletal pre-Lie 2-algebras and strict pre-Lie 2-algebras in detail.

Let $(A_0,A_1,\dM=0,\cdot,l_3)$ be a  skeletal pre-Lie 2-algebra. Condition $(b_1)$ in Definition \ref{defi:2-term pre} implies that $(A_0,\cdot)$ is a pre-Lie algebra. Define $\rho$ and $\mu$ from $A_0$ to $\gl(A_1)$ by
\begin{equation}
  \rho(u)m=u\cdot m,\quad \mu(u)m=m\cdot u,\quad \forall ~u\in A_0,~m\in A_1.
\end{equation}
Condition $(b_2)$ and $(b_3)$  in Definition \ref{defi:2-term pre} implies that $(A_1;\rho,\mu)$ is a representation of the pre-Lie algebra $(A_0,\cdot)$. Furthermore, Condition (c) exactly means that $l_3$ is a 3-cocycle on $A_0$ with values in $A_1$. Summarize the discussion above, we have

\begin{thm}\label{thm:classify}
There is a one-to-one correspondence between   skeletal pre-Lie $2$-algebras and triples $((A_0,\cdot),(A_1;\rho,\mu),l_3)$, where $(A_0,\cdot)$ is a pre-Lie algebra, $(A_1;\rho,\mu)$ is a representation of $(A_0,\cdot)$, and $l_3$ is a $3$-cocycle on $(A_0,\cdot)$ with values in $A_1$.
\end{thm}

Recall that a skew-symmetric bilinear form $\omega:\wedge^2A\longrightarrow A$ on a pre-Lie algebra $(A,\cdot_A)$ is called {\bf invariant} if
\begin{equation}\label{eq:invariant}
\omega(u\cdot_A v-v\cdot_Au,w)+\omega(v, u\cdot_A w)=0, \quad \forall u,v,w\in A.
\end{equation}
Equivalently, $\omega([u, v]_A,w)+\omega(v, u\cdot_A w)=0,$ where $[\cdot,\cdot]_A$ is the Lie bracket in the sub-adjacent Lie algebra of $A$.
\begin{lem}
  Let $\omega$ be a skew-symmetric invariant bilinear form on a pre-Lie algebra $(A,\cdot_A)$. Then we have
  \begin{equation}\label{eq:invariantnew}
    \omega(u\cdot_A v,w)=\omega(u,w\cdot_A v).
  \end{equation}
\end{lem}
\pf By \eqref{eq:invariant}, we have
\begin{equation}\label{eq:invariant1}
  \omega(u\cdot_Aw-w\cdot_Au,v)+\omega(w,u\cdot_Av)=0.
\end{equation}
Since $\omega$ is skew-symmetric, by \eqref{eq:invariant} and \eqref{eq:invariant1}, we have
$$
-\omega(w\cdot_Au,v)-\omega(v\cdot_Au,w)=0,
$$
which implies that $\omega(u\cdot_A v,w)=\omega(u,w\cdot_A v).$ \qed\vspace{3mm}

Define $\varphi:\wedge^2A\otimes A\longrightarrow\mathbb R$ by
\begin{equation}\label{eq:3cocyclel3}
  \varphi(u,v,w)=\omega(u\cdot_A v-v\cdot_A u,w).
\end{equation}
\begin{pro}\label{pro:closed}
Let $\omega$ be a skew-symmetric invariant bilinear form on a pre-Lie algebra $(A,\cdot_A)$. Then $\varphi$ defined by \eqref{eq:3cocyclel3} is a $3$-cocycle on $A$ with values in $\mathbb R$, i.e. $d\varphi=0$.
\end{pro}
\pf For any $u,v,w,p\in A$, by \eqref{eq:invariant} and \eqref{eq:invariantnew}, we have
\begin{eqnarray*}
  d\varphi(u,v,w,p)&=&-\varphi(v,w,u\cdot_A p)+\varphi(u,w,v\cdot_A p)-\varphi(u,v,w\cdot_A p)\\
  &&-\varphi([u,v]_A,w,p)+\varphi([u,w]_A,v,p)-\varphi([v,w]_A,u,p)\\
  &=&-\omega([v,w]_A,u\cdot_A p)+\omega([u,w]_A,v\cdot_A p)-\omega([u,v]_A,w\cdot_A p)\\
  &&-\omega([[u,v]_A,w]_A+c.p.,p)\\
  &=&\omega(w,v\cdot_A(u\cdot_A p))-\omega(w,u\cdot_A(v\cdot_A p))+\omega(w,[u,v]_A\cdot_A p)\\
  &=&\omega(w,v\cdot_A(u\cdot_A p)-u\cdot_A(v\cdot_A p)+(u\cdot_Av)\cdot_Ap-(v\cdot_Au)\cdot_Ap)\\
  &=&0,
\end{eqnarray*}
which finishes the proof. \qed
\begin{ex}{\rm
  Let $\omega$ be a skew-symmetric invariant bilinear form on a pre-Lie algebra $(A,\cdot_A)$. Consider the graded vector space $\huaA=A_0\oplus A_1$ where $A_0=A, A_1=\mathbb R$. Define $\dM:\mathbb R\longrightarrow A,$ $\cdot:A_i\otimes A_j\longrightarrow A_{i+j}$, $0\leq i+j\leq 1$, and $l_3:\otimes A_0\longrightarrow A_1$ by
  \begin{eqnarray*}
    \dM&=&0,\\
    u\cdot v&=&u\cdot_A v,\\
    u\cdot m&=&m\cdot u=0,\\
    l_3(u,v,w)&=&\varphi(u,v,w),
  \end{eqnarray*}
  for any $u,v,w\in A$ and $m\in A_1.$ By Proposition \ref{pro:closed}, it is straightforward to verify that $\huaA=(A,\mathbb R,\dM=0,\cdot,l_3=\varphi)$ is a pre-Lie 2-algebra.  Furthermore, $\omega$ ia a closed 2-form on the Lie algebra $\g(A)$, i.e.
  $$
  \omega([u,v]_A,w)+\omega([v,w]_A,u)+\omega([w,u]_A,v)=0,
  $$
 which implies that $l_3(u,v,w)+l_3(v,w,w)+l_3(w,u,v)=0.$  Thus, the skeletal Lie 2-algebra $\huaG(\huaA)$ is also strict.
  }
\end{ex}

Now we turn to the study on   strict pre-Lie 2-algebras. First we introduce the notion of crossed modules of pre-Lie algebras, which could give rise to crossed modules of Lie algebras.
\begin{defi}
 A {\bf crossed module of pre-Lie algebras} is a quadruple $((A_0,\cdot_0),(A_1,\cdot_1),\dM,(\rho,\mu))$ where $(A_0,\cdot_0)$ and $(A_1,\cdot_1)$ are pre-Lie algebras, $\dM:A_1\longrightarrow A_0$ is a homomorphism of pre-Lie algebras, and $(\rho,\mu)$ is an action of $(A_0,\cdot_0)$ on $A_1$ such that for all $u\in A_0$ and $m,n\in A_1$, the following equalities are satisfied:
 \begin{itemize}
      \item[\rm (C1)] $\dM(\rho(u)m)=u\cdot_0\dM m,\quad \dM(\mu(u)m)=(\dM m)\cdot_0u,$
   \item[\rm (C2)] $\rho(\dM m)n=\mu(\dM n)m= m\cdot_1n.$
 \end{itemize}
\end{defi}
\begin{ex}{\rm
Let $(A,\cdot_A)$ be a pre-Lie algebra and $B\subset A$ an ideal. Then it is straightforward to see that $((A,\cdot),(B,\cdot|_B),\inc,(\rho,\mu))$ is a crossed module of pre-Lie algebras, where $\inc$ is the inclusion, and $(\rho,\mu)$ are given by $\rho(u)v=u\cdot_Av,~\mu(u)v=v\cdot_A u,$ for all $u\in A,v\in B.$
}\end{ex}

\begin{pro}
Let $((A_0,\cdot_0),(A_1,\cdot_1),\dM,(\rho,\mu))$ be a crossed module of pre-Lie algebras. Then we have
 \begin{eqnarray}
 \label{eq:crossed1}\rho(u)(m\cdot_1 n)&=&(\rho(u)m)\cdot_1 n+m\cdot_1\rho(u)n-(\mu(u)m)\cdot_1 n,\\
  \label{eq:crossed2} \mu(u)(m\cdot_1n)&=&\mu(u)(n\cdot_1m)+m\cdot_1\mu(u)n-n\cdot_1\mu(u)m.
   \end{eqnarray}
   Consequently, there is a pre-Lie algebra structure $\cdot$ on the direct sum $A_0\oplus A_1$ given by
   \begin{equation}
     (u+m)\cdot (v+n)=u\cdot_0v+\rho(u)n+\mu(v)m+m\cdot_1n.
   \end{equation}
\end{pro}
\pf Since $(\rho,\mu)$ is an action of $A_0$ on $A_1$, we have
\begin{eqnarray*}
  \rho(u)\rho(\dM m)n&=&\rho(u\cdot_0\dM m)n-\rho(\dM m\cdot_0u)n+\rho(\dM m)\rho(u)n\\
  &=&\rho(\dM \rho(u)m)n-\rho(\dM \mu(u)m)n+\rho(\dM m)\rho(u)n.
\end{eqnarray*}
The second equality is due to  (C1). By (C2), we obtain \eqref{eq:crossed1}.  \eqref{eq:crossed2} can be obtained similarly. The other conclusion is obvious. \qed

\begin{thm}\label{thm:one-to-one}
  There is a one-to-one correspondence between   strict pre-Lie $2$-algebras and crossed modules of pre-Lie algebras.
\end{thm}
\pf
Let $(A_0, A_1,\dM,\cdot,l_3=0)$ be a   strict pre-Lie 2-algebra. We construct a crossed module of pre-Lie algebras as follows. Obviously, $(A_0,\cdot)$ is a pre-Lie algebra. Define a multiplication $\cdot_1$ on $A_1$ by
\begin{equation}
  m\cdot_1 n=(\dM m)\cdot n=m\cdot \dM n.
\end{equation}
Then by Conditions $(a_1)$ and $(b_2)$ in Definition \ref{defi:2-term pre}, we have
\begin{eqnarray*}
  && m\cdot_1 (n\cdot_1p)-(m\cdot_1n)\cdot_1p-n\cdot_1 (m\cdot_1p)+(n\cdot_1 m)\cdot_1p\\
  &=&(\dM m)\cdot ((\dM n)\cdot p)-\dM((\dM m)\cdot n)\cdot p-(\dM n)\cdot ((\dM m)\cdot p)+\dM((\dM n)\cdot m)\cdot_1p\\
  &=&(\dM m)\cdot ((\dM n)\cdot p)-((\dM m)\cdot \dM n)\cdot p-(\dM n)\cdot ((\dM m)\cdot p)+((\dM n)\cdot \dM m)\cdot_1p=0,
\end{eqnarray*}
which implies that $(A_1,\cdot_1)$ is a pre-Lie algebra. Also by Condition $(a_1)$, we deduce that $\dM$ is a homomorphism between pre-Lie algebras. Define $\rho,\mu:A_0\longrightarrow\gl(A_1)$ by
\begin{equation}
  \rho(u)m=u\cdot m,\quad \mu(u)m=m\cdot u.
\end{equation}
By Conditions $(b_2)$ and $(b_3)$ in Definition \ref{defi:2-term pre}, it is straightforward to deduce that $(\rho,\mu)$ is an action of $(A_0,\cdot)$ on $A_1$. By Conditions $(a_1)$ and $(a_2)$, we deduce that Condition (C1) hold. Condition (C2) follows from the definition of $\cdot_1$ directly. Thus, $((A_0,\cdot),(A_1,\cdot_1),\dM,(\rho,\mu))$ constructed above is a crossed module of pre-Lie algebras.

Conversely, a crossed module of pre-Lie algebras  $((A_0,\cdot),(A_1,\cdot_1),\dM,(\rho,\mu))$  gives rise to a   strict pre-Lie 2-algebra $(A_0, A_1,\dM,\cdot,l_3=0)$, where $\cdot:A_i\otimes A_j\longrightarrow A_{i+j},~0\leq i+j\leq 1$ is given by
$$
u\cdot v=u\cdot_1 v,\quad u\cdot m=\rho(u)m,\quad m\cdot u=\mu(u)m.
$$
The crossed module conditions give various conditions for a
strict pre-Lie 2-algebra. We omit details.\qed\vspace{3mm}

A pre-Lie algebra has its sub-adjacent Lie algebra. Similarly, a crossed module of pre-Lie algebras has its sub-adjacent crossed module of Lie algebras. Recall that {\bf a crossed module of Lie algebras} is a quadruple
$(\frkh_1,\frkh_0,dt,\phi)$, where
$\frkh_1$ and $\frkh_0$ are Lie algebras,
$dt:\frkh_1\longrightarrow\frkh_0$ is a Lie algebra homomorphism and
$\phi:\frkh_0\longrightarrow\Der(\frkh_1)$ is an action of Lie
algebra $\frkh_0$ on Lie algebra $\frkh_1$ as a derivation, such
that
$$
dt(\phi_X(A))=[X,dt(A)]_{\frkh_0},\quad \phi_{dt(A)}(B)=[A,B]_{\frkh_1},\quad \forall X\in \frkh_0, A,B\in\frkh_1.
$$
\begin{pro}
  Let $((A_0,\cdot_0),(A_1,\cdot_1),\dM,(\rho,\mu))$ be a crossed module of pre-Lie algebras and $\g(A_0),\g(A_1)$ the corresponding sub-adjacent Lie algebras of $(A_0,\cdot_0),(A_1,\cdot_1)$ respectively. Then $(\g(A_0),\g(A_1),\dM,\rho-\mu)$ is a crossed module of Lie algebras.
\end{pro}
\pf The fact that $\dM$ is a homomorphism between pre-Lie algebras implies that $\dM$ is also a homomorphism between Lie algebras. Since $(A_1;\rho,\mu)$ is a representation of $(A_0,\cdot_0)$, $(A_1;\rho-\mu)$ is a representation of the Lie algebra $\g(A_0)$. By (C1), we have $\dM((\rho-\mu)(u)m)=[u,\dM m]_0$. By (C2), we have $(\rho-\mu)(\dM m)n=[m,n]_1$. Thus, $(\g(A_0),\g(A_1),\dM,\rho-\mu)$ is a crossed module of Lie algebras.\qed

\section{Categorification of $\mathcal{O}$-operators}

Let $\huaG=(\g_0,\g_{1}, \frkd, \frkl_2,\frkl_3)$ be a Lie 2-algebra and $(\rho_0,\rho_1,\rho_2)$ be a representation of $\huaG$ on a 2-term complex of vector spaces $\huaV=V_1\stackrel{\dM}{\longrightarrow}V_0$.
\begin{defi}\label{defi:O-operator}
  A triple $(T_0,T_1,T_2)$, where $T_0:V_0\longrightarrow \g_0,T_1:V_1\longrightarrow \g_1$ is a chain map, and $T_2:\wedge^2V_0\longrightarrow\g_1$ is a linear map, is called an  $\mathcal{O}$-operator on $\huaG$ associated to the representation  $(\rho_0,\rho_1,\rho_2)$, if for all $u,v,v_i\in V_0$ and $m\in V_1$ the following conditions are satisfied:
  \begin{itemize}
    \item[\rm(i)] $T_0\big(\rho_0(T_0u)v-\rho_0(T_0v)u\big)-\frkl_2(T_0u,T_0v)=\dM T_2(u,v);$

    \item[\rm(ii)] $T_1\big(\rho_1(T_1m)v-\rho_0(T_0v)m\big)-\frkl_2(T_1m,T_0v)= T_2(\dM m,v);$

    \item[\rm(iii)] \begin{eqnarray*}
          &&\frkl_2(T_0(v_1),T_2(v_2,v_3))+T_2\big(v_3,\rho_0(T_0v_1)v_2-\rho_0(T_0v_2)v_1\big)\\
          &&+T_1\big(\rho_1(T_2(v_2,v_3))v_1+\rho_2(T_0v_2,T_0v_3)v_1\big)+c.p.+\frkl_3(T_0v_1,T_0v_2,T_0v_3)=0.
        \end{eqnarray*}
    \end{itemize}
    \end{defi}

\begin{ex}{\rm
  Let $\huaA=(A_0,A_1,\dM,\cdot,l_3)$ be a pre-Lie 2-algebra. Then, $(T_0=\id_{A_0},T_1=\id_{A_1},T_2=0)$ is an $\mathcal{O}$-operator on the Lie 2-algebra $\huaG(\huaA)$ associated to the representation  $(L_0,L_1,L_2)$ given in Theorem \ref{thm:main1}.}
\end{ex}

    Define a degree $0$ multiplication $\cdot:V_i\otimes V_j\longrightarrow V_{i+j}$, $0\leq i+j\leq 1$, on $\huaV$ by
    \begin{equation}\label{eq:formularm}
      u\cdot v=\rho_0(T_0u)v,\quad u\cdot m=\rho_0(T_0u)m,\quad m\cdot u=\rho_1(T_1m)u.
    \end{equation}
    Define $l_3: \wedge^2V_0\otimes V_0\longrightarrow V_1$ by
    \begin{equation}\label{eq:formularl3}
      l_3(v_1,v_2,v_3)=-\rho_1(T_2(v_1,v_2))v_3-\rho_2(T_0v_1,T_0v_2)v_3.
    \end{equation}
Now, Condition (iii) in Definition \ref{defi:O-operator} reads
\begin{eqnarray}
\nonumber&&\frkl_2(T_0(v_1),T_2(v_2,v_3))+T_2(v_3,v_1\cdot v_2-v_2\cdot v_1)-T_1(l_3(v_1,v_2,v_3))+c.p.\\
\label{eq:Ocon}&&+\frkl_3(T_0v_1,T_0v_2,T_0v_3)=0.
\end{eqnarray}
    \begin{thm}
      Let $(\rho_0,\rho_1,\rho_2)$ be a representation of $\huaG$ on $\huaV$ and $(T_0,T_1,T_2)$ an  $\mathcal{O}$-operator on $\huaG$ associated to the representation  $(\rho_0,\rho_1,\rho_2)$. Then, $(V_0,V_1,\dM,\cdot,l_3)$ is a pre-Lie  $2$-algebra, where the multiplication ``$\cdot$'' and $l_3$ are given by \eqref{eq:formularm} and \eqref{eq:formularl3} respectively.
    \end{thm}
\pf
By the fact that $\dM\circ \rho(x)=\rho(x)\circ\dM$ for all $x\in\g_0$, we deduce that
$$
\dM(u\cdot m)=\dM\rho_0(T_0u)m=\rho_0(T_0u)\dM m=u\cdot \dM m.
$$
By the fact that both $(T_0,T_1)$ and $(\rho_0,\rho_1)$ are chain maps, we have
$$
\dM(m\cdot u)=\dM(\rho_1(T_1m)u)=\delta(\rho_1(T_1m))u=\rho_0(\frkd T_1m)u=\rho_0(T_0\dM m)u=(\dM m)\cdot u.
$$
Similarly, we have
$$
(\dM m)\cdot n=\rho_0(T_0\dM m)n=\rho_0(\frkd T_1 m)n=\delta(\rho_1(T_1 m))n=\rho_1(T_1 m)(\dM n)=m\cdot(\dM n).
$$
Thus, Conditions $(a_1)$-$(a_3)$ in Definition \ref{defi:2-term pre} hold.
For all $u,v,w\in A_0$, we  have
    \begin{eqnarray*}
     &&u\cdot(v\cdot w)-(u\cdot v)\cdot w-v\cdot(u\cdot w)+(v\cdot u)\cdot w\\
    &=&\rho_0(T_0u)\rho_0(T_0v)w-\rho_0(T_0(\rho_0(T_0u)v))w-\rho_0(T_0v)\rho_0(T_0u)w+\rho_0(T_0(\rho_0(T_0v)u))w\\
    &=&[\rho_0(T_0u),\rho_0(T_0v)]w-\rho_0\Big(T_0(\rho_0(T_0u)v)-T_0(\rho_0(T_0v)u)\Big)w\\
    &=&\rho_0(\frkl_2(T_0u,T_0v))w-\dM\rho_2(T_0u,T_0v)w-\rho_0\Big(T_0(\rho_0(T_0u)v)-T_0(\rho_0(T_0v)u)\Big)w\\
    &=&-\rho_0(\dM T_2(u,v))w-\dM\rho_2(T_0u,T_0v)w\\
    &=&-\dM \rho_1(T_2(u,v))w-\dM\rho_2(T_0u,T_0v)w\\
    &=&\dM l_3(u,v,w),
    \end{eqnarray*}
which implies that Condition $(b_1)$ in Definition \ref{defi:2-term pre} holds. Similarly, Conditions $(b_2)$ and $(b_3)$ also hold.

    The left hand side of Condition (c) is equal to
    \begin{eqnarray*}
      &&\rho_0(T_0v_0)l_3(v_1,v_2,v_3)-\rho_0(T_0v_1)l_3(v_0,v_2,v_3)+\rho_0(T_0v_2)l_3(v_0,v_1,v_3)\\
      &&+\rho_1(T_1l_3(v_1,v_2,v_0))v_3-\rho_1(T_1l_3(v_0,v_2,v_1))v_3+\rho_1(T_1l_3(v_0,v_1,v_2))v_3\\
      &&+\rho_1(T_2(v_1,v_2))(v_0\cdot v_3)+\rho_2(T_0v_1,T_0v_2)(v_0\cdot v_3)-\rho_1(T_2(v_0,v_2))(v_1\cdot v_3)\\
      &&-\rho_2(T_0v_0,T_0v_2)(v_1\cdot v_3)+\rho_1(T_2(v_0,v_1))(v_2\cdot v_3)+\rho_2(T_0v_0,T_0v_1)(v_2\cdot v_3)\\
      &&+\rho_1(T_2(\rho_0(T_0v_0)v_1-\rho_0(T_0v_1)v_0,v_2))v_3+\rho_2(T_0(\rho_0(T_0v_0)v_1-\rho_0(T_0v_1)v_0),T_0v_2)v_3\\
      &&-\rho_1(T_2(\rho_0(T_0v_0)v_2-\rho_0(T_0v_2)v_0,v_1))v_3-\rho_2(T_0(\rho_0(T_0v_0)v_2-\rho_0(T_0v_2)v_0),T_0v_1)v_3\\
      &&+\rho_1(T_2(\rho_0(T_0v_1)v_2-\rho_0(T_0v_2)v_1,v_0))v_3+\rho_2(T_0(\rho_0(T_0v_1)v_2-\rho_0(T_0v_2)v_1),T_0v_0)v_3\\
      &=&-\rho_0(T_0v_0)\rho_1(T_2(v_1,v_2))v_3-\rho_0(T_0v_0)\rho_2(T_0v_1,T_0v_2)v_3\\
      &&+\rho_0(T_0v_1)\rho_1(T_2(v_0,v_2))v_3+\rho_0(T_0v_1)\rho_2(T_0v_0,T_0v_2)v_3\\
      &&-\rho_0(T_0v_2)\rho_1(T_2(v_0,v_1))v_3-\rho_0(T_0v_2)\rho_2(T_0v_0,T_0v_1)v_3\\
     && +\rho_1(T_1l_3(v_1,v_2,v_0))v_3-\rho_1(T_1l_3(v_0,v_2,v_1))v_3+\rho_1(T_1l_3(v_0,v_1,v_2))v_3\\
     &&+\rho_1(T_2(v_1,v_2))\rho_0(v_0)v_3+\rho_2(T_0v_1,T_0v_2)\rho_0(v_0) v_3-\rho_1(T_2(v_0,v_2))\rho_0(v_1) v_3\\
      &&-\rho_2(T_0v_0,T_0v_2)\rho_0(v_1) v_3+\rho_1(T_2(v_0,v_1))\rho_0(v_2) v_3+\rho_2(T_0v_0,T_0v_1)\rho_0(v_2) v_3\\
     &&+\rho_1(T_2(\rho_0(T_0v_0)v_1-\rho_0(T_0v_1)v_0,v_2))v_3+\rho_2(T_0(\rho_0(T_0v_0)v_1-\rho_0(T_0v_1)v_0),T_0v_2)v_3\\
      &&-\rho_1(T_2(\rho_0(T_0v_0)v_2-\rho_0(T_0v_2)v_0,v_1))v_3-\rho_2(T_0(\rho_0(T_0v_0)v_2-\rho_0(T_0v_2)v_0),T_0v_1)v_3\\
      &&+\rho_1(T_2(\rho_0(T_0v_1)v_2-\rho_0(T_0v_2)v_1,v_0))v_3+\rho_2(T_0(\rho_0(T_0v_1)v_2-\rho_0(T_0v_2)v_1),T_0v_0)v_3\\
      &=&\Big(-[\rho_0(T_0v_0),\rho_1(T_2(v_1,v_2))]+c.p.\Big)v_3+\Big(-[\rho_0(T_0v_0),\rho_2(T_0v_1,T_0v_2)]+c.p.\Big)v_3\\
      &&+\Big(\rho_1(T_1l_3(v_0,v_1,v_2))+c.p.\Big)v_3+\Big(\rho_1T_2(v_0\cdot v_1-v_1\cdot v_0,v_2)+c.p.\Big)v_3\\
      &&+\Big(\rho_2(T_0(v_0\cdot v_1-v_1\cdot v_0),T_0v_2)+c.p.\Big)v_3\\
      &=&\Big(-\rho_1\frkl_2(T_0v_0,T_2(v_1,v_2))+\rho_2(T_0v_0,\dM T_2(v_1,v_2))+c.p.\Big)v_3\\&&+\Big(-[\rho_0(T_0v_0),\rho_2(T_0v_1,T_0v_2)]+c.p.\Big)v_3
      +\Big(\rho_1(T_1l_3(v_0,v_1,v_2))+c.p.\Big)v_3\\&&+\Big(\rho_1T_2(v_0\cdot v_1-v_1\cdot v_0,v_2)+c.p.\Big)v_3
      +\Big(\rho_2(T_0(v_0\cdot v_1-v_1\cdot v_0),T_0v_2)+c.p.\Big)v_3.
    \end{eqnarray*}
By Condition (ii) in Definition \ref{defi:O-operator}, we have
$$
\rho_2(T_0v_0,\dM T_2(v_1,v_2))+c.p.+\rho_2(T_0(v_0\cdot v_1-v_1\cdot v_0),T_0v_2)+c.p.=\rho_2(\frkl_2(T_0v_0,T_0v_1),T_0v_2)+c.p..
$$
By the fact that $(\rho_0,\rho_1,\rho_2)$ is a representation, we have
$$
[\rho_0(T_0v_1),\rho_2(T_0v_2,T_0v_3)]+c.p.-\rho_2(\frkl_2(T_0v_1,T_0v_2),T_0v_3)+c.p.=\rho_1\frkl_3(T_0v_1,T_0v_2,T_0v_3).
$$
By \eqref{eq:Ocon}, we deduce that Condition (c) in Definition \ref{defi:2-term pre} holds. Thus, $(V_0,V_1,\dM,\cdot,l_3)$ is a pre-Lie 2-algebra.
This finishes the proof.\qed

\begin{cor}
Let $(\rho_0,\rho_1,\rho_2)$ be a representation of Lie $2$-algebra $\huaG$ on $\huaV$ and $(T_0,T_1,T_2)$ an  $\mathcal{O}$-operator on $\huaG$ associated to the representation  $(\rho_0,\rho_1,\rho_2)$. Then,  $(T_0,T_1,T_2)$ is a homomorphism from the Lie $2$-algebra  $\huaG(\huaV)$ to $\huaG$.
\end{cor}

\section{Solutions of 2-graded Classical Yang-Baxter Equations}

Let $\huaG=(\g_0,\g_{1}, \frkd, \frkl_2)$ be a
 strict Lie 2-algebra and $r\in \g_0 \otimes
\g_{1} \oplus  \g_{1} \otimes \g_0$ and $\frkr \in
\g_{1} \otimes \g_{1}$. Denote $R=r-(\dM\otimes1+1\otimes\dM)\frkr$. Then, the classical Yang-Baxter
equation for $R$ in the semidirect product Lie algebra $\g_0\ltimes \g_1=(\g_0\oplus \g_1,[\cdot,\cdot]_s)$ together with $(\frkd\otimes1-1\otimes \frkd )R=0$ are called  the {\bf 2-graded classical Yang-Baxter Equations} (2-graded CYBE) in the strict Lie $2$-algebra $\huaG$, where $[\cdot,\cdot]_s$ is the semidirect product Lie algebra structure given by \eqref{eq:semidirect}. More precisely, the 2-graded CYBE reads:
\begin{itemize}
\item[\rm(a)] $R$ is skew-symmetric,
\item[\rm(b)] $[R_{12},R_{13}]_s+[R_{13},R_{23}]_s+[R_{12},R_{23}]_s=0$,
\item[\rm(c)] $(\frkd\otimes1-1\otimes \frkd )r=0$.
\end{itemize}
For $R=\sum\limits_i a_i\otimes b_i$,
\begin{equation}
R_{12}=\sum\limits_i a_i\otimes b_i\otimes 1;\quad R_{13}=\sum\limits_i
a_i\otimes1\otimes b_i;\quad R_{23}=\sum\limits_i 1\otimes
a_i\otimes b_i.
\end{equation}
Let $(\rho_0,\rho_1)$ be a strict representation of the Lie 2-algebra $\huaG=(\g_0,\g_{1}, \frkd, \frkl_2)$ on the 2-term complex of vector space $\huaV:V_1\stackrel{\dM}{\longrightarrow}V_0$.
We view $\rho_0\oplus \rho_1$ a linear map from $\g_0\oplus \g_1$ to $ \gl(V_0\oplus V_1)$ by
\begin{equation}
  (\rho_0\oplus\rho_1)(x+a)(u+m)=\rho_0(x)(u)+\rho_0(x)m+\rho_1(a)u.
\end{equation}
By straightforward computations, we have

\begin{lem}\label{lem:Ooperatorequi}
With the above notations,  $\rho_0\oplus \rho_1:\g_0\oplus \g_1\longrightarrow \gl(V_0\oplus V_1)$ is a representation of $(\g_0\oplus \g_1,[\cdot,\cdot]_s)$ on $V_0\oplus V_1$. Furthermore, $(T_0,T_1)$ is an $\mathcal O$-operator on $\huaG$ associated to the representation $(\rho_0,\rho_1)$ if and only if \begin{itemize}
  \item[\rm(a)] $T_0+ T_1:V_0\oplus V_1\longrightarrow \g_0\oplus \g_1$ is an $\mathcal O$-operator on the Lie algebra $(\g_0\oplus \g_1,[\cdot,\cdot]_s)$ associated to the representation $\rho_0\oplus \rho_1$,
       \item[\rm(b)] $T_0\circ \dM=\frkd\circ T_1.$
 \end{itemize}
\end{lem}

Let $(\rho^*_0,\rho^*_1)$ be the dual representation of $(\rho_0,\rho_1)$. Then we have the semidirect product Lie 2-algebra $\bar{\huaG}=\huaG\ltimes_{(\rho_0^*,\rho_1^*)}\huaV^*$, where $\bar{\huaG}_0=\g_0\oplus V_1^*,~\bar{\huaG}_1=\g_1\oplus V_0^*,$ and $\bar{\frkd}=\frkd\oplus \dM^*$. It is obvious that
$$
\overline{T_0}+\overline{T_1}\in V_0^*\otimes \g_0\oplus V_1^*\otimes \g_1 \in (\bar{\huaG}_1\otimes \bar{\huaG}_0)\oplus (\bar{\huaG}_0\otimes \bar{\huaG}_1),
$$
where $\overline{T_0}$ and $\overline{T_1}$ are given by \eqref{eq:Tbar}.
\begin{thm}\label{thm:solution}
  Let $(\rho_0,\rho_1)$ be a strict representation of the Lie $2$-algebra $\huaG=(\g_0,\g_{1}, \frkd, \frkl_2)$ on the $2$-term complex of vector space $\huaV:V_1\stackrel{\dM}{\longrightarrow}V_0$, and $T_0:V_0\longrightarrow\g_0,T_1:V_1\longrightarrow\g_1$ linear maps. Then, $(T_0,T_1)$ is an $\mathcal O$-operator on the Lie $2$-algebra $\huaG$ associated to the representation $(\rho_0,\rho_1)$ if and only if $\overline{T_0}+\overline{T_1}-\sigma(\overline{T_0}+\overline{T_1})$ is a solution of the $2$-graded CYBE in the semidirect product Lie $2$-algebra $\bar{\huaG}$.
\end{thm}
\pf It is obvious that $(\rho_0\oplus \rho_1)^*=\rho_0^*\oplus\rho_1^*:\g_0\oplus\g_1\longrightarrow\gl(V_1^*\oplus V_0^*)$. By Lemma \ref{lem:Ooperatorequi} and Theorem \ref{thm:O-operator}, $(T_0,T_1)$ is an $\mathcal O$-operator on the Lie 2-algebra $\huaG$ if and only if $\overline{T_0}+\overline{T_1}-\sigma(\overline{T_0}+\overline{T_1})$ is a solution of the CYBE in the semidirect product Lie algebra $(\g_0\ltimes \g_1)\ltimes_{\rho_0^*\oplus\rho_1^*}(V_1^*\oplus V_0^*)$, and $T_0\circ \dM=\frkd\circ T_1.$ Note that the semidirect product Lie algebra $(\g_0\ltimes \g_1)\ltimes_{\rho_0^*\oplus\rho_1^*}(V_1^*\oplus V_0^*)$ is exactly the same as the semidirect product Lie algebra $\bar{\huaG}_0\ltimes \bar{\huaG_1}$. Furthermore, $T_0\circ \dM=\frkd\circ T_1$ if and only if $(\bar{\frkd}\otimes1-1\otimes \bar{\frkd} )(\overline{T_0}+\overline{T_1})=0$. Thus, $(T_0,T_1)$ is an $\mathcal O$-operator on the Lie $2$-algebra $\huaG$ associated to the representation $(\rho_0,\rho_1)$ if and only if $\overline{T_0}+\overline{T_1}-\sigma(\overline{T_0}+\overline{T_1})$ is a solution of the $2$-graded CYBE in the semidirect product Lie $2$-algebra $\bar{\huaG}$.\qed\vspace{3mm}

Let $\huaA=(A_0,A_1,\dM,\cdot)$ be a strict  pre-Lie 2-algebra. Then, $\huaG(\huaA)=(A_0,A_1,\dM,\frkl_2)$ is a strict Lie $2$-algebra, where $\frkl_2$ is given by \eqref{eq:l21} and \eqref{eq:l22}. Furthermore, $(L_0,L_1)$ is a strict representation of the Lie $2$-algebra $\huaG(\huaA)$ on the complex of vector spaces $A_1\stackrel{\dM}{\longrightarrow} A_0$, where $L_0,L_1$ are given by \eqref{eq:L0} and \eqref{eq:L1} respectively. Let $\{e_i\}_{1 \leq i\leq  k}$ and $\{\frke_j\}_{1 \leq j\leq  l}$ be the basis of $A_0$ and $A_1$ respectively, and denote by $\{e_i^*\}_{1 \leq i\leq  k}$ and $\{\frke_j^*\}_{1 \leq j\leq  l}$ the dual basis.

\begin{thm}\label{thm:solutionpreLie2}
  With the above notations,
  \begin{equation}
    R=\sum_{i=1}^k (e_i\otimes e_i^*-e_i^*\otimes
e_i)+\sum_{j=1}^l (\frke_j\otimes \frke_j^*-\frke_j^*\otimes
\frke_j)
  \end{equation}
  is a solution of the $2$-graded CYBE in the strict Lie $2$-algebra $\huaG(\huaA)\ltimes_{(L_0^*,L_1^*)}\huaA^*$.
\end{thm}
\pf It is obvious that $(T_0=\id_{A_0},T_1=\id_{A_1})$ is an $\mathcal O$-operator on $\huaG(\huaA)$ associated to the representation $(L_0,L_1)$. By Theorem \ref{thm:solution}, $$\overline{T_0}+\overline{T_1}-\sigma(\overline{T_0}+\overline{T_1})=\sum_{i=1}^k (e_i\otimes e_i^*-e_i^*\otimes
e_i)+\sum_{j=1}^l (\frke_j\otimes \frke_j^*-\frke_j^*\otimes
\frke_j)$$ is a solution of the $2$-graded CYBE in the strict Lie $2$-algebra $\huaG(\huaA)\ltimes_{(L_0^*,L_1^*)}\huaA^*$.\qed\vspace{3mm}

At the end of this section, we consider the construction of strict Lie 2-bialgebras in \cite[Proposition  4.4]{BaiShengZhu}. In fact, there are pre-Lie 2-algebras behind the construction.

Let $(A,\cdot_A)$ be a pre-Lie algebra. Then $(A;L,R)$ is a representation of $(A,\cdot_A)$. Furthermore, $(A^*;L^*-R^*,-R^*)$ is also a representation of $(A,\cdot_A)$. Let $A_0=A$ and $A_1=A^*$. Define a multiplication $\cdot:A_i\otimes A_j\longrightarrow A_{i+j},~0\leq i+j\leq 1,$ by
\begin{equation}\label{eq:multiplication}
  x\cdot y=x\cdot_A y,\quad x\cdot \xi=\ad^*_x\xi,\quad \xi\cdot x=-R^*_x\xi,\quad\forall x,y\in A,~\xi\in A^*.
\end{equation}

On the other hand, consider its sub-adjacent Lie algebra $\g(A)$. Define a skew-symmetric operation $\frkl_2:A_i\wedge A_j\longrightarrow A_{i+j},~0\leq i+j\leq 1,$ by
\begin{equation}\label{eq:l2AA*}
  \frkl_2(x,y)=[x,y]_A=x\cdot_A y-y\cdot_A x,\quad \frkl_2(x,\xi)=-\frkl_2(\xi,x)=L^*_x\xi.
\end{equation}

\begin{pro}\label{pro:pre-Lie 2 and Lie 2}
  Let $(A,\cdot_A)$ be a pre-Lie algebra, and $\dM:A^*\longrightarrow A$ a linear map. If $(A,A^*,\dM,\cdot) $ is a pre-Lie $2$-algebra, $(\g(A),A^*,\dM,\frkl_2)$ is a Lie $2$-algebra, where $\cdot$ and $\frkl_2$ are given by \eqref{eq:multiplication} and \eqref{eq:l2AA*} respectively.

  Conversely, if $(\g(A),A^*,\dM,\frkl_2)$ is a Lie $2$-algebra, in which $\dM:A^*\longrightarrow A$ is skew-symmetric, $(A,A^*,\dM,\cdot) $ is a pre-Lie $2$-algebra.
\end{pro}
\pf If $(A,A^*,\dM,\cdot) $ is a pre-Lie $2$-algebra, we have
$$
\dM (\ad_x^*\eta)=x\cdot\dM \eta,\quad \dM(-R_y^*\xi)=(\dM\xi)\cdot y,\quad \ad_{\dM\xi}^*\eta=-R^*_{\dM \eta}\xi,\quad\forall x,y\in A,~\xi,\eta\in A^*.
$$
Therefore, we have
\begin{eqnarray*}
  \dM \frkl_2(x,\eta)&=&\dM L_x^*\eta=\ad_x^*\eta+R_x^*\eta=x\cdot\dM\eta-(\dM\eta)\cdot x=\frkl_2(x,\dM\eta),\\
  \frkl_2(\dM\xi,\eta)&=&L_{\dM\xi}^*\eta=\ad_{\dM\xi}^*\eta+R_{\dM\xi}^*\eta=\ad_{\dM\xi}^*\eta-\ad_{\dM\eta}^*\xi=\frkl_2(\xi,\dM\eta).
\end{eqnarray*}
Since $L^*$ is a representation of the Lie algebra $\g(A)$ on $A^*$, it is obvious that the other conditions in the definition of a Lie 2-algebra are also satisfied. Thus, $(\g(A),A^*,\dM,\frkl_2)$ is a Lie $2$-algebra.

Conversely, if $(\g(A),A^*,\dM,\frkl_2)$ is a Lie $2$-algebra, we have
$$
\dM \frkl_2(x,\eta)=\frkl_2(x,\dM\eta),\quad\frkl_2(\dM\xi,\eta)=\frkl_2(\xi,\dM\eta),
$$
which implies that
$$
\dM L^*_x\eta=L_x\dM\eta-R_x\dM\eta,\quad L_{\dM\xi}^*\eta=-L_{\dM\eta}^*\xi.
$$
If $\dM$ is skew-symmetric, we can obtain
\begin{eqnarray*}
  \langle\dM R_x^*\eta,\xi\rangle&=& \langle R_x^*\eta,-\dM \xi\rangle=\langle \eta,R_x\dM \xi\rangle\\
  &=&\langle \eta,L_x\dM\xi-\dM L_x^*\xi\rangle=\langle\dM L_x^*\eta-L_x\dM\eta,\xi\rangle\\
  &=&\langle-R_x\dM\eta,\xi\rangle,
\end{eqnarray*}
which implies that
\begin{equation}\label{eq:ttt1}
\dM (\eta\cdot x)=(\dM\eta)\cdot x.
\end{equation}
Furthermore, we have
$$
\dM (\ad_x^*\eta)=\dM(L_x^*\eta-R_x^*\eta)=L_x\dM\eta,
$$
which implies that
\begin{equation}\label{eq:ttt2}
\dM(x\cdot \eta)=x\cdot \dM \eta.
\end{equation}
Also by the fact that $\dM$ is skew-symmetric, we have
\begin{eqnarray*}
  \langle\ad^*_{\dM \xi}\eta,x\rangle&=&\langle\eta,L_x\dM\xi-R_x\dM\xi\rangle=\langle\dM L^*_x\eta-\dM R^*_x\eta,\xi\rangle\\
  &=&\langle L_x\dM\eta-R_x\dM\eta+R_x\dM\eta,\xi\rangle\\
  &=&\langle R_{\dM\eta}x,\xi\rangle=\langle x,-R^*_{\dM\eta}\xi\rangle,
\end{eqnarray*}
which implies that  $\ad^*_{\dM \xi}\eta=-R^*_{\dM\eta}\xi$, i.e.
\begin{equation}\label{eq:ttt3}
(\dM\xi)\cdot\eta=\xi\cdot(\dM\eta).
\end{equation}
By \eqref{eq:ttt1}-\eqref{eq:ttt3}, we deduce that Conditions $(a_1)$-$(a_3)$ in Definition \ref{defi:2-term pre} hold. It is obvious that the other conditions also hold. Thus, $(A,A^*,\dM,\cdot) $ is a pre-Lie $2$-algebra. \qed\vspace{3mm}

By Proposition \ref{pro:pre-Lie 2 and Lie 2} and Proposition 4.4 in \cite{BaiShengZhu}, we have
\begin{cor}
   Let $(A,A^*,\dM,\cdot) $ be a pre-Lie $2$-algebra, where $\cdot$ is given by \eqref{eq:multiplication} and $\dM$ is skew-symmetric. Then $r$ given by \eqref{eq:rrr} is a solution of the $2$-graded CYBE in the strict Lie $2$-algebra $(\g(A),A^*,\dM,\frkl_2)$, where $\frkl_2$ is given by \eqref{eq:l2AA*}.
\end{cor}

\end{document}